\documentclass[showpacs,preprint]{revtex4}

\usepackage{epsfig}
\usepackage{amsmath}

\begin{document}

\title{Joint use of the Weniger transformation and hyperasymptotics 
for accurate asymptotic
evaluations of a class of saddle-point integrals. II. Higher-order 
transformations}

\author{Riccardo Borghi}

\affiliation{Dipartimento di Elettronica Applicata, Universit\`a 
degli Studi ``Roma Tre"\\
Via della Vasca Navale 84, I-00146 Rome, Italy}

\begin{abstract}
The use of hyperasymptotics and the Weniger transformation has been 
proposed, in a 
joint fashion, for decoding the divergent asymptotic series generated 
by 
the steepest descent on a wide class of saddle-point integrals 
{evaluated across Stokes sets}
[R. Borghi, Phys. Rev. E {\bf 78,} 026703 (2008)]. In the present 
sequel, 
the full development of the H-WT up to the second order in H is 
derived. 
Numerical experiments, carried out on several classes of saddle-point 
integrals, 
including the swallowtail diffraction catastrophe, show the 
effectiveness of the 
2nd-level H-WT, in particular when the integrals are evaluated beyond 
the asymptotic realm.
\end{abstract}


\maketitle

\section{Introduction}
\label{intro}

The present paper is a sequel of a previous work\cite{borghiPRE-08}
concerning 
the evaluation of saddle-point integrals of the form
\begin{equation}
\mathcal{I}(k)=\displaystyle\int_{\mathcal{C}}\,
g(s)\,\exp[-k\,f(s)]\,\mathrm{d}s,
\label{sd.1}
\end{equation}
where $\mathcal{C}$ is a suitable integration path in the complex 
$s$-plane,
$g(s)$ and $f(s)$ are functions which, for simplicity, will be 
assumed to be nonsingular,
and $k$ will be intended as a ``large'' (in modulus) complex 
parameter.
As is well known, the numerical evaluation of integrals 
of the kind in Eq. (\ref{sd.1}) is customarily required for solving 
several
classes of physical problems, occurring in optics, quantum mechanics, 
statistical physics, 
fluid mechanics, and so on.
In optics, the evaluation of several diffraction integrals is 
customarily
carried out asymptotically by identifying the parameter $k$ as the 
wavenumber 
of the radiation\cite{born&wolf}. In quantum mechanics, the same role 
is played by
the inverse of the Planck's constant, while in fluid mechanics by the 
Reynold's number\cite{berryLectures-89}.

In the stationary phase treatment of diffraction integrals the values 
of the
associated complex wavefield are asymptotically evaluated by taking 
the contributions 
coming from the stationary points of $f(s)$, each of them associated 
to a ``ray" 
in the corresponding geometrical picture. 
Of particular importance is the birth and the death, as the spatial 
parameters ruling the ``phase" function $\mathrm{i} f(s)$ vary, of 
``evanescent" rays across sets of 
codimension 1, named ``Stokes sets" \cite{wrightJPA-80,berryNL-90}.

The $\delta$-, or Weniger, 
transformation\cite{wenigerCPR-89,wenigerJMP-04,calicetiArXiv07} (WT 
for short),
is particularly efficient for resumming the factorial divergent 
asymptotic series well away from Stokes sets, as well as sets where 
two or more saddles are 
symmetrically placed in the complex singulant space\cite{endnote31}.
{
Unfortunately, as with other resummation techniques\cite{wenigerCPR-89,brezinski},  
the WT fails to perform across Stokes sets.
The reason of such a failure stems to the extreme "specialization" of 
the transformation itself, which requires, for a successful 
resummation, an alternating sign pattern of the sequence of 
the single terms of the series\cite{jentschuraCPC-99}.
Several methods have been conceived for resumming nonalternating, 
slowly convergent or divergent, sequences 
\cite{jentschuraPRD-00,calicetiArXiv07}, some of them being based
on the serial combination of various resummation 
techniques\cite{jentschuraCPC-99,aksenovCPC-03}.
For the class of saddle-point integrals in Eq. (\ref{sd.1}), the 
marriage between hyperasymptotics\cite{berryPRSA-90,berryPRSA-91} (H 
for short) and the WT \cite{borghiPRE-08},
generating the so-called H-WT (which stands for 
hyperasymptotic-Weniger transformation),
allows the WT to successfully operate also across Stokes sets.
}

Basically, the H-WT consists in the sequential application, to the 
integral in Eq. (\ref{sd.1}), of a classical hyperasymptotic 
treatment, as described
in Ref. \cite{berryPRSA-91}, followed by the action of the WT on all  
asymptotic divergent series generated by H. 
In particular, the results obtained have shown how the 
1st-order H-WT, for which only the first-stage of H is anticipated to 
the WT, is able to provide relative errors several orders of 
magnitude smaller than those achievable via the use of full 
hyperasymptotic treatments and with considerably lighter 
computational complexity and effort. A key aspect is that, 
differently from H, the first truncation operated on the starting 
asymptotic series has not to be an optimal, in the sense of 
superasymptotics (i.e., at the least term) one, but rather the 
corresponding 
truncation order, say $N$, must be used as a free parameter for the 
subsequent application of the WT. 

A question which was not addressed in Ref. \cite{borghiPRE-08}, but 
mentioned only in the last sentence, is whether WT and H can be 
combined to higher orders in H, and if so, how the accuracy improves 
with order. The present paper is aimed at giving a first answer to 
such a question. We shall limit our analysis only to the second stage 
of H. Further increasing of the H-WT order would be achievable along 
the same guidelines outlined here. On the other hand, it should be 
noted how, on increasing the order of H, the number of asymptotic 
series associated to the corresponding remainder that have to be 
resummed exponentially grows for topologies involving more than two 
saddles
and, at the same time, the number of free parameters (i.e., the 
truncation orders at each H step) linearly increases. Accordingly, 
from a mere computational viewpoint, it is mandatory to find a 
compromise between the H-WT order and the computational effort. Some 
limits of the 1st-order H-WT have already been emphasized in Ref. 
\cite{borghiPRE-08}, where asymptotic evaluations of saddle-point 
integrals for ``small" values of the asymptotic parameter were 
considered. In such cases, in fact, to increase the parameter $N$ of 
the 1st-order H-WT does not necessarily lead to an improvement of the 
reached accuracy, but often results in the opposite, i.e., a 
worsening of it. It is just the above scenario that we are interested 
in when the 
2nd-level H-WT will be developed. Numerical experiments will be 
carried 
out on the class of saddle-point integrals already considered in the 
numerical 
sections of the first paper \cite{borghiPRE-08}. Moreover, asymptotic 
evaluations of the so-called swallowtail diffraction catastrophe 
\cite{berryPIO-80} will be proposed as a new numerical experiment. 
The swallowtail function is defined via Eq. (\ref{sd.1}) with $f(s)$ 
being a 5th-order polynomial with respect to $s$, thus involving a 
four saddle network. We will present a study of the accuracy 
achievable via H-WT asymptotic evaluations of the swallowtail 
diffraction catastrophe 
for points placed at the Stokes set, following the prescriptions by 
Berry and Howls\cite{berryNL-90}. In doing so, we will find that the 
corresponding asymptotic expanding coefficients can be expressed in 
closed-form terms. 

{For any practical implementation of the H-WT, a key role is
played by the numerical evaluation of the corresponding 
hyperterminants\cite{berryPRSA-90,berryPRSA-91} which are defined 
through
suitable multiple integrals. For the lowest-order hyperterminant the 
exact 
analytical expression is available from 
literature\cite{berryPRSA-90}, but
unfortunately this is not true for higher-order hyperterminants,
included those involved in the 2nd-level H-WT.
In the present paper we solve the problem of the 2nd-level 
hyperterminant exact evaluation for a particular, but very important, 
choice 
of the hyperterminant parameters, which often occurs in the 
implementation of H for evaluating a wide class of saddle-point 
integrals. Up to our knowledge, this is a new result
which also provides an interesting connection of such 
hypeterminants to the Meijer-G functions \cite{GR}. 
Moreover, although the closed-form evaluation of 2nd-level 
hypeterminants for arbitrary choices of their parameters 
seems to remain an open problem, in the present paper we find a 
semi-analytical representation which turns out to be 
suitable for numerical calculations via standard integration 
packages.  
}
Similarly as done in Ref. \cite{borghiPRE-08}, 
one of our aims is to keep the paper reasonably self-consistent. 
Accordingly, in the next section a brief review of H, up to the 
2nd-level,
is given. As far as the WT is concerned, we believe that what is 
contained
in Ref. \cite{borghiPRE-08}, together with
the extensive bibliography, should be enough also for a nonexpert 
reader.
For this reason, we do not repeat it in the present paper.

\section{Resuming Hyperasymptotics}
\label{hyperAsymptotics}

\subsection{Preliminaries and notations}
\label{notations}

For simplicity, we shall refer to the asymptotic evaluation of 
saddle-point integrals of 
the type in Eq. (\ref{sd.1}) where the set of saddle points of 
$f(s)$ will be denoted $\mathcal{S}$ and the integration path 
$\mathcal{C}$ will be thought of as the union of a finite number of 
steepest descent arcs each of them, say $\mathcal{C}_n$, 
passing through the contributive saddle point $s_n$, which will be 
supposed to be a
simple one. Accordingly, the quantity $\mathcal{I}(k)$ can generally  
be written as
\begin{equation}	
\mathcal{I}(k)=\displaystyle\int_{\mathcal{C}}\,g(s)\,\exp[-k\,f(s)]\,\mathrm{d}s=
\displaystyle\sum_{n \in \mathcal{S}'}\,\mathcal{I}^{(n)}(k),
\label{sdReview.5}
\end{equation}
where  $\mathcal{S}'$ denotes the subset of $\mathcal{S}$ containing 
all the contributive saddles, and
\begin{equation}
	\mathcal{I}^{(n)}(k)=
	\displaystyle\int_{\mathcal{C}_n}\, g(s)\,\exp[-k f(s)]\,\mathrm{d}s.
	\label{sdReview.5.1}
\end{equation}
The last integral can be written as\cite{berryPRSA-91}
\begin{equation}
\mathcal{I}_n(k)=k^{-1/2}\,\exp(-kf_n)\,T^{(n)}(k),
\label{sdReview.3}
\end{equation}
where $f_n=f(s_n)$, and where $T^{(n)}(k)$ can \emph{formally} be 
written through the following asymptotic series expansion:
\begin{equation}
T^{(n)}(k)=\displaystyle\sum_{r=0}^{\infty}\,k^{-r}\,T^{(n)}_r,
\label{sdReview.4}
\end{equation}
the expanding coefficients $T^{(n)}_r$ being expressed via the
integral representation \cite{berryPRSA-91}
\begin{equation}
T^{(n)}_r=
\displaystyle\frac{(r-1/2)!}{2\pi\mathrm{i}}\,
\oint_{n}\,\displaystyle\frac{g(s)}{[f(s)-f_n]^{r+1/2}}\,\mathrm{d}s,
\label{sdReview.5.0}
\end{equation}
where the subscript $n$ denotes a small positive loop around the 
saddle $s_n$.

\subsection{Development of H up to the second stage}
\label{H}

H starts by writing Eq. (\ref{sdReview.4}) in the form
\begin{equation}
T^{(n)}(k)=\displaystyle\sum_{r=0}^{N-1}\,k^{-r}\,T^{(n)}_r+ 
R^{(n)}(k,N),
\label{sdReview.5bis}
\end{equation}
where $N$ represents a positive integer and
$R^{(n)}(k,N)=\sum_{r=N}^\infty\,k^{-r}\,T^{(n)}_r$, 
denotes the corresponding remainder
which, due to the diverging character of the asymptotic series, turns 
out to be a \emph{diverging} quantity too.
H is based on a couple of fundamental results, found via a nontrivial 
analysis in
Ref. \cite{berryPRSA-91}. The first is that the value of the 
expanding coefficients
$T^{(n)}_r$ at the saddle $s_n$ is strictly related to the values of 
the expanding coefficients $T^{(m)}_r$ at all those saddles, say 
$\{s_m\}$, 
which are \emph{adjacent} to $s_n$, via the following formal
{resurgence} relation\cite{berryPRSA-91}:
\begin{equation}
T^{(n)}_r=\displaystyle\frac 1{2\pi\mathrm{i}}\,
\displaystyle\sum_{m\in \mathcal{A}_n}\,(-1)^{\gamma_{nm}}\,
\displaystyle\sum_{l=0}^\infty\,
\displaystyle\frac{(r-l-1)!}{F^{r-l}_{nm}}\,T^{(m)}_l,
\label{resurgence.1}
\end{equation}
where $\mathcal{A}_n$ denotes the set containing the indexes 
pertinent to 
all saddles adjacent to $s_n$, the quantities $F_{nm}$, called 
\emph{singulants}, 
are defined by
\begin{equation}
F_{nm}=f_m-f_n,
\label{singulant}
\end{equation}
and the binary quantities $\gamma_{nm} \in \{0,1\}$  are obtained 
through a topological rule\cite{berryPRSA-91}.
The other fundamental tool  of H is the  following 
integral representation of the 
remainder $R^{(n)}(k,N)$\cite{berryPRSA-91}:
\begin{equation}
\begin{array}{l}
R^{(n)}(k,N)=\displaystyle\frac 1{2\pi\mathrm{i}}\,
\displaystyle\sum_{m \in 
\mathcal{A}_n}\,\displaystyle\frac{(-1)^{\gamma_{nm}}}{(kF_{nm})^N}\\
\\
\times
\displaystyle\int_0^\infty\,
\mathrm{d}v\,\displaystyle\frac{v^{N-1}\,\exp(-v)}{1-\displaystyle\frac 
v{kF_{nm}}}\,
T^{(m)}\left(\displaystyle\frac v{F_{nm}}\right).
\end{array}
\label{remainder.2}
\end{equation}
Equations (\ref{sdReview.4})-(\ref{remainder.2}) allow 
hyperasymptotic 
expansions for the saddle integral in Eq. (\ref{sdReview.5.1}) to be 
built up 
in principle to any order\cite{berryPRSA-91}.
For instance, the direct substitution of Eq. (\ref{sdReview.4}) into 
Eq. (\ref{remainder.2})
leads to the 1st-stage hyperasymptotic expansion \cite{borghiPRE-08},
\begin{equation}
\begin{array}{lcl}
T^{(n)}(k)  =  
\displaystyle\sum_{r=0}^{N-1}\,k^{-r}\,T^{(n)}_r\\
\\
+
\displaystyle\frac {(-1)^N}{2\pi\mathrm{i}}\,
\displaystyle\sum_{m \in \mathcal{A}_n}\,(-1)^{\gamma_{nm}}\\
\\
\,\,\,\,\,\,\,\,\,\,\,\,\,\,\,\,\,\,\,\,\,\times
\displaystyle\sum_{r=0}^{\infty}\,
(-1)^r\,k^{-r}\,T^{(m)}_r\,K^{(1)}_{N-r}(-kF_{nm}),
\end{array}
\label{Istage.complete}
\end{equation}
where the function $K^{(1)}_{n}(\beta)$, called \emph{hyperterminant} 
of 
order 1\cite{berryPRSA-90,berryPRSA-91}, is defined through the 
integral
\begin{equation}
\begin{array}{lcl}
K^{(1)}_{n}(\beta)=
\displaystyle\frac 1{\beta^n}\,
\displaystyle\int_0^\infty\,
\mathrm{d}v\,\displaystyle\frac{v^{n-1}\,\exp(-v)}{1+\displaystyle\frac 
v\beta},
\end{array}
\label{remainder.3.1}
\end{equation}
where, in order for it to converge, $n>0$.
Moreover, it can be shown that\cite{borghiPRE-08}
\begin{equation}
K^{(1)}_{n}(\beta)=
\exp(\beta)\,\displaystyle\frac{E_n(\beta)}{\beta^{n-1}}\,(n-1)!+(-1)^{n-1}\,\mathrm{i}\pi 
\epsilon\,\exp(\beta),
\label{remainder.3.1.1}
\end{equation}
where $E_n(\cdot)$ denotes the exponential integral function 
\cite{GR}, while $\epsilon$ equals 1 if $\beta<0$ and zero otherwise.
The presence of the term containing $\epsilon$ has to be ascribed to 
the evaluation of the integral in Eq. (\ref{remainder.3.1}), when 
$\beta<0$,
in the Cauchy principal-value sense. Equation (\ref{Istage.complete}) 
represents
the first hyperasymptotic stage, at which the divergence of the 
asymptotic series 
in Eq. (\ref{sdReview.4}) is led back to the presence of adjacent 
saddles \cite{berryPRSA-91}. Furthermore, the asymptotic series in 
Eq. (\ref{Istage.complete}) are only formal, since for $r>N$ the 
terminant $K^{(1)}_{N-r}$ diverges.
In Ref. \cite{borghiPRE-08} Eq. (\ref{Istage.complete}) was taken as
the starting point for introducing the H-WT. 
In particular, instead of using the WT directly on the single terms 
of 
the series in Eq. (\ref{sdReview.4}), it is employed for resumming 
the asymptotic series associated
to all saddles $s_m$, with $m\in\mathcal{A}_n$, which appear in 
Eq. (\ref{Istage.complete}).
Of course, due to the fact that $r \le N$ in Eq. 
(\ref{Istage.complete}),
it is mandatory that $N$ be left as a free parameter, in order for 
the WT 
to be able in decoding the above asymptotic series.
The 2nd-level H can be derived  by truncating each of the asymptotic 
series 
in Eq. (\ref{Istage.complete}) at an order, say $M$, and by 
generating, for \emph{each}
adjacent saddle $s_m$, with $m\in \mathcal{A}_n$, a list of 
asymptotic series 
associated to all saddles, say $s_h$, such that $h \in \mathcal{A}_m$.
In Appendix \ref{IIstage}, only for the reader convenience, the 
derivation
of the 2nd-level hyperasymptotic expansion of the integral in Eq. 
(\ref{sd.1})
is briefly recalled according to the formalism of 
Ref.~\cite{berryPRSA-91}. In particular, it is found that 
\begin{equation}
\begin{array}{lcl}
T^{(n)}(k)=\displaystyle\sum_{r=0}^{N-1}\,k^{-r}\,T^{(n)}_r\\
\\+
\displaystyle\frac {(-1)^N}{2\pi\mathrm{i}}\,
\displaystyle\sum_{m \in \mathcal{A}_n}\,(-1)^{\gamma_{nm}}\\
\\
\,\,\,\,\,\,\,\,\,\,\,\,\,\,\,\,\,\,\,\,\,\times
\displaystyle\sum_{r=0}^{M-1}\,
(-1)^r\,k^{-r}\,T^{(m)}_r\,K^{(1)}_{N-r}(-kF_{nm})\\
\\
+
\displaystyle\frac{(-1)^{N+M}}{(2\pi\mathrm{i})^2}\,
\displaystyle\sum_{m \in \mathcal{A}_n}
\displaystyle\sum_{h \in 
\mathcal{A}_m}\,(-1)^{\gamma_{nm}+\gamma_{mh}}\,\\
\\
\,\,\,\,\,\,\,\,\,\,\,\,\,\,\,\,\,\,\,\,\,\,\,\,\,\,\,\,\,\,\times
\displaystyle\sum_{r=0}^\infty\,k^{-r}T^{(h)}_r\,
K^{(2)}_{M-r,N-M}\left(-kF_{nm};-\displaystyle\frac{F_{mh}}{F_{nm}}\right),

\end{array}
\label{IIstage.complete}
\end{equation}
where $K^{(2)}_{n,m}(\beta;\gamma)$, the hyperterminant of order 2, 
is now defined through the double integral
\begin{equation}
\begin{array}{lcl}
K^{(2)}_{n,m}(\beta;\gamma) = 
\displaystyle\frac 1{\beta^{n+m}}\,\\
\\
\times
\displaystyle\int_0^\infty\,
\displaystyle\int_0^\infty\,
\mathrm{d}u\,\mathrm{d}v\,
\exp(-u-\gamma\,v)\,
\displaystyle\frac{u^{m-1} v^{n-1}}{\left(1+\displaystyle\frac 
u\beta\right)\,\left(1+\displaystyle\frac v u \right)}.
\end{array}
\label{terminant2}
\end{equation}
Similarly as we done in Ref. \cite{borghiPRE-08},
Eq. (\ref{IIstage.complete}) can be used to give estimates
of $T^{(n)}(k)$, as functions of the two (free) parameters 
$N$ and $M$, by resumming, via the WT, all asymptotic series
generated at the second stage of H which are inside the double 
sum with respect to $h$ and $m$.

\section{On the evaluation of $K^{(2)}_{n,m}(\beta;\gamma)$}
\label{K2}

{The numerical evaluation of the hyperterminants
represents a step of fundamental importance for any practical
implementation of the H-WT algorithm. Unfortunately, 
differently from the lowest-order H-WT, for which the 
corresponding hyperterminants are achievable via the closed-form
expression in Eq. (\ref{remainder.3.1.1}), there are no analytical 
expressions available for higher-order hyperterminants. 
In a series of important papers,  
Olde Daalhuis~\cite{daalhiusJCAM-96,daalhiusJCAM-98}
addressed the general problem of the hyperterminants  evaluation,
up to arbitrary precisions, through the use of convergent series 
representations based on hypergeometric functions.
However, for the particular case of the 2nd-level 
hyperterminant, it seems that some new, at least up to our knowledge, 
results could be  established.

From Eq. (\ref{terminant2}) where, in order for it to converge, $n>0$ 
and $m>0$,
the hyperterminant can be written as
\begin{equation}
\begin{array}{lcl}
K^{(2)}_{n,m}(\beta;\gamma) = \displaystyle\frac 1{\beta^{n+m}}\,\\
&&\\
\times
\displaystyle\int_0^\infty\,
\displaystyle\int_0^\infty\,
\mathrm{d}u\,\mathrm{d}v\,
\exp(-u-\gamma\,v)\,
\displaystyle\frac{u^{m} v^{n-1}}{\left(1+\displaystyle\frac 
u\beta\right)\,(u+v)},
\end{array}
\label{K2.0.1}
\end{equation}
and, by formally expanding the factor $1/(1+u/\beta)$ as
a geometric series, after some algebra takes the form
\begin{equation}
\begin{array}{lcl}
K^{(2)}_{n,m}(\beta;\gamma) = \displaystyle\frac {(-1)^{m}}{\beta^{n}}
\displaystyle\sum_{k=m}^\infty\,
\left(-\displaystyle\frac 1\beta\right)^k\,\\
\\
\times
\displaystyle\int_0^\infty\,
\displaystyle\int_0^\infty\,
\mathrm{d}u\,\mathrm{d}v\,
\exp(-u-\gamma\,v)\,
\displaystyle\frac{u^{k} v^{n-1}}{u+v}=\\
\\
=(-1)^m\,
\displaystyle\frac{(n-1)!}{(\beta\gamma)^{n}}\\
\\
\times
\displaystyle\sum_{k=m}^\infty\,
\left(-\displaystyle\frac 1\beta\right)^k\,
\displaystyle\frac{k!}{k+n}\,
F\left(n,1;k+n+1;1-\displaystyle\frac{1}{\gamma}\right),
\end{array}
\label{K2.0.2}
\end{equation}
where $F(\cdot,\cdot;\cdot;\cdot)$ denotes the hypergeometric 
function\cite{GR}.
Although the series in Eq. (\ref{K2.0.2}) is divergent, it can be 
decoded 
via Borel summation, i.e., by replacing the term $k!$ by its integral 
representation, i.e.,
\begin{equation}
k!=\displaystyle\int_0^\infty\,\mathrm{d}t\,
\exp(-t)\,t^k,
\label{K2.0.3}
\end{equation}
which, once substituted into Eq.~(\ref{K2.0.2}), leads to
\begin{equation}
\begin{array}{lcl}
K^{(2)}_{n,m}(\beta;\gamma) = (-1)^m\,
\displaystyle\frac{(n-1)!}{(\beta\gamma)^{n}}\,
\displaystyle\int_{0}^{\infty}\,
\mathrm{d}t\,
\exp(-t)\\
\\
\times
\displaystyle\sum_{k=m}^{\infty}\,
\left(-\displaystyle\frac t\beta\right)^k\,
\displaystyle\frac{1}{k+n}\,
F\left(n,1;k+n+1;1-\displaystyle\frac{1}{\gamma}\right).
\end{array}
\label{K2.0.3.1}
\end{equation}
Although, as we shall in a moment, it is possible to express the 
series
inside the last equation through a closed form, it is better to carry 
out  
the evaluations for the case $\gamma=1$ and $\gamma \ne 1$ separately.

On letting into Eq.~(\ref{K2.0.3.1}) $\gamma=1$ we have
\begin{equation}
\begin{array}{l}
K^{(2)}_{n,m}(\beta;1) = 
(-1)^m\,\displaystyle\frac{(n-1)!}{\beta^{n}}\\
\\
\times
\displaystyle\int_{0}^{\infty}\,
\mathrm{d}t\,
\exp(-t)\,
\displaystyle\sum_{k=m}^{\infty}\,
\left(-\displaystyle\frac t\beta\right)^k\,
\displaystyle\frac{1}{k+n}=\\
\\
=
\displaystyle\frac{(n-1)!}{(m+n)\beta^{n-1}}\\
\\
\times
\displaystyle\int_0^\infty\,
\mathrm{d}t\,
\exp(-t)\,
\left(\displaystyle\frac t\beta \right)^m\,
F\left(m+n,1;m+n+1;-\displaystyle\frac t\beta\right).
\end{array}
\label{K2.0.3.2}
\end{equation}
The integral in Eq.~(\ref{K2.0.3.2}) can be evaluated 
by using the representation of the hypergeometric function 
given by formula 9.34.7 of Ref. \cite{GR}. In particular, it turns 
out that
\begin{equation}
\begin{array}{l}
K^{(2)}_{n,m}(\beta;1)=
\displaystyle\frac{(n-1)!}{\beta^{n-1}}\,\\
\\
\times
\displaystyle\int_0^\infty\,
\mathrm{d}t\,
\exp(- t)\,
\left(\displaystyle\frac t\beta \right)^{m+1}\,
G^{12}_{22}
\left(\displaystyle\frac t\beta\left|\begin{array}{l} 
-n-m,\,-1\\-1,\,-n-m-1\end{array}\right.\right),
\end{array}
\label{hyper2.10.3}
\end{equation}
where $G^{mn}_{pq}(\cdot)$ denotes the Meijer function\cite{GR}.
Finally, by using formulas 9.31.5, 
7.813.1, and 9.31.2 of \cite{GR}, after some algebra it is found that
\begin{equation}
    \begin{array}{lcl}
    K^{(2)}_{n,m}(\beta;1)& = &(n-1)!\,
    G^{31}_{23}\left(\beta\left|\begin{array}{c}
    1-n-m,1\\1-n,1-n-m,0\end{array}\right.\right).
\end{array}
\label{terminant2.1}
\end{equation}
Equation~(\ref{terminant2.1})  represents one of the main results of 
the present paper. As we shall see in the numerical section, in 
applying the 2nd-level H-WT  the evaluation of the 
hyperterminants $K^{(2)}_{n,m}(\beta;\gamma)$ is often required 
for $\gamma=1$. This happens whenever the contributive saddle
$s_{n}$ turns out to be adjacent to itself after two hyperasymptotic
stages, i.e., when $h=n$ into Eq. 
(\ref{IIstage.complete})\cite{berryPRSA-91}. 

For $\gamma \ne 1$, the series inside the integral in 
Eq.~(\ref{K2.0.2})
can still be expressed through a closed form, although the subsequent 
integral
unfortunately not. However, a semi-analytical expression, which turns 
out to be suitable for being evaluated
via standard numerical integration packages can be derived.
In doing this,  Eq.~(\ref{K2.0.2}) is first  rewritten as 
\begin{equation}
\begin{array}{lcl}
K^{(2)}_{n,m}(\beta;\gamma) =
(-1)^{m}\,\displaystyle\frac{(n-1)!}{(\beta\gamma)^{n}}\\
\\
\times
\left[
\mathcal{S}-
\displaystyle\sum_{k=0}^{m-1}\,
\left(-\displaystyle\frac 1\beta\right)^k\,
\displaystyle\frac{k!}{k+n}\,
F\left(n,1;k+n+1;\displaystyle\frac{\gamma-1}{\gamma}\right)
\right],
\end{array}
\label{gn1.1}
\end{equation}
where
\begin{equation}
\begin{array}{lcl}
\mathcal{S}=
\displaystyle\sum_{k=0}^\infty\,
\left(-\displaystyle\frac 1\beta\right)^k\,
\displaystyle\frac{k!}{k+n}\,
F\left(n,1;k+n+1;\displaystyle\frac{\gamma-1}{\gamma}\right),
\end{array}
\label{gn1.2}
\end{equation}
so that the task is to evaluate the series in Eq.~(\ref{gn1.2}) for 
$\gamma \ne 1$. 
On substituting from Eq.~(\ref{K2.0.3}) into Eq.~(\ref{gn1.2}) we have
\begin{equation}
\begin{array}{lcl}
\mathcal{S}=
\displaystyle\int_{0}^{\infty}\,\mathrm{d}t\,
\exp(-t)\,\\
\\
\times
\displaystyle\sum_{k=0}^\infty\,
\left(-\displaystyle\frac t\beta\right)^k\,
\displaystyle\frac{F\left(n,1;k+n+1;1-\displaystyle\frac{1}{\gamma}\right)}{k+n}=\\

\\
=\gamma\,\displaystyle\int_{0}^{\infty}\,\mathrm{d}t\,
\exp(-t)\,\\
\\
\times
\displaystyle\sum_{k=0}^\infty\,
\left(-\displaystyle\frac t\beta\right)^k\,
\displaystyle\frac{F(1+k,1;n+1+k;1-\gamma)}{k+n},
\end{array}
\label{gn1.3}
\end{equation}
where use has been made of the relation[see Ref.~\cite{PrudnikovIII} 
p.~347]
\begin{equation}
\begin{array}{lcl}
F\left(n,1;k+n+1;1-\displaystyle\frac{1}{\gamma}\right)
=\gamma\,F(1+k,1;n+1+k;1-\gamma).
\end{array}
\label{gn1.4}
\end{equation}
Finally, on writing Eq.~(\ref{gn1.3}) as
\begin{equation}
\begin{array}{lcl}
\mathcal{S}=\displaystyle\frac\gamma n\,
\displaystyle\int_{0}^{\infty}\,\mathrm{d}t\,\exp(-t)\,\\
\\
\times
\displaystyle\sum_{k=0}^\infty\,
\displaystyle\frac{(1)_{k}}{k!}\,
\displaystyle\frac{(n)_{k}}{(n+1)_{k}}\,
\left(-\displaystyle\frac t\beta\right)^k\,
F(1+k,1;n+1+k;1-\gamma),
\end{array}
\label{gn1.5}
\end{equation}
where $(\cdot)_{k}$ denotes the Pochhammer symbol,
formula 6.7.1.8 of Ref.~\cite{PrudnikovIII} gives at once
\begin{equation}
\begin{array}{lcl}
\mathcal{S}=\displaystyle\frac\gamma n\,
\displaystyle\int_{0}^{\infty}\,\mathrm{d}t\,
\displaystyle\frac{\exp(-t)}{1+\displaystyle\frac t\beta}\,
F\left(1,1;n+1;1-\displaystyle\frac\gamma{1+\displaystyle\frac 
t\beta}\right).
\end{array}
\label{gn1.6}
\end{equation}
Notice that, when $\mathrm{Re}[\beta]>0$, the function $S$
can also be evaluated through the alternative form
\begin{equation}
\begin{array}{lcl}
\mathcal{S}=\displaystyle\frac{\beta\gamma}n\,
\displaystyle\int_{0}^{1}\,\mathrm{d}p\,
\displaystyle\frac
{\exp\left[-\beta\left(\displaystyle\frac 1p-1\right)\right]}{p}\,
F(1,1;n+1;1-\gamma p).
\end{array}
\label{gn1.7}
\end{equation}
Although it seems that the above expressions cannot be further 
simplified, the numerical evaluation of the function $S$ can be done 
with high accuracies by using standard integration packages. 
Finally, it should be stressed that, for $\beta<0$,
the  evaluation of the double integral
defining $K^{(2)}_{n,m}(\beta;\gamma)$ has to be done, with respect 
to the $v$-variable, in the Cauchy principal value sense, in order to
overcome the singularity placed at $v=-\beta$. This, in turn, implies 
that an extra term must be added to the result. In 
Appendix~\ref{appC} such term is analytically 
evaluated starting from the definition in Eq.~(\ref{terminant2}), and
turns out to be
\begin{equation}
    \begin{array}{l}
\mathrm{i}\pi\,(-1)^{n+m-1}(n-1)!\displaystyle\frac{\exp[\beta(1-\gamma)]}
{(-\beta)^{n-1}}
	    \,E_{n}(-\beta\gamma).
	\end{array}
\label{extraTerm}
\end{equation}
\\
All subsequent numerical experiments will be done 
within the \emph{Mathematica} language. 
}

\section{Numerical experiments}
\label{numericalExperiments}

\subsection{Evaluation of the Airy function 
across the Stokes line}
\label{airy}

{
Consider the evaluation of the Airy function, defined as
\begin{equation}
\mathrm{Ai}(x)=\displaystyle\frac1{2\pi}\,
\displaystyle\int_{\mathcal{C}}\,
\exp\left[\mathrm{i}\left(\displaystyle\frac{s^3}3+xs\right)\right]\,
\mathrm{d}s,
\label{airy.1}
\end{equation}
which is of the form given in Eq. (\ref{sd.1}) with $g(s)=1/(2\pi)$, 
$f(s)=-\mathrm{i} (s^3/3+xs)$, and $k=1$. 
The detailed analysis of the saddle topology, as well as the
expanding coefficients $T^{(n)}_r$ has been summarized in Ref. 
\cite{borghiPRE-08},
so that here it will not be given again. We only recall that the two 
saddles
are $s_1= (-x)^{1/2}$ and $s_2=-s_1$, and that $\mathcal{A}_{{1\atop 
2}}=\{{2\atop 1}\}$, $\gamma_{12}=0$, $\gamma_{21}=1$. 
We focus our attention on the evaluation of the Airy function across 
the Stokes line\cite{boyd-99}, i.e., for $\arg\{x\}=2\pi/3$, in order 
to compare the performances
of the 2nd-level H-WT with respect those displayed by the 1st-order 
H-WT in the same situation. More precisely, we write the argument of 
the Airy function as 
$x=(3/4\times F)^{2/3}\,\exp(\mathrm{i}2\pi/3)$, where $F$ is a real 
positive parameter, whose value coincides with the singulant $F_{12}$.
The study of the asymptotic evaluation of the Airy function across 
its Stokes line
has played a pivotal role in the development of several asymptotic 
technique,
mainly in light of the relative simplicity of the involved saddle 
topology. Such
a simplicity could help in grasping, whereas possible, some 
conceptual aspects related
to the use of the H-WT.
Differently from what done in Ref.~\cite{borghiPRE-08},
where the relative error values were displayed via the use of tables, 
in the present paper we are going to resort to graphical visualizations,
due to the presence of the two ``free'' parameters $N$ and $M$.
In the first experiment, whose results are shown in
Figure~\ref{FigErrorAiryF16},  
the Airy function is evaluated for $F=16$. 
Note that the same experiment was carried out in 
Ref.~\cite{berryPRSA-90} via the use of H.
The values of the relative error, obtained through the 1st-level 
H-WT, are shown, as black dots, versus the values of $N$, reported on 
the abscissa axis. For each value of $N$, the values of the 
relative error obtained via the 2nd-level H-WT, with $M \in [3,N-1]$, 
are also plotted and, for the sake of clarity, are joined with lines 
of different color each of them, which  corresponds to a different value of $N$, 
departing from $N$ itself.
This can be noted from the figure, where it is immediately seen how 
the higher the $N$,
the longer the corresponding coloured ``leg'' is. 
From a first look to the figure, it appears that the relative error,
obtained with both the 1st- and the 2nd-level H-WT, is lower bounded.
We shall find that all subsequent numerical experiments present the 
same 
characteristic. As a first remark, it should be noting that
the improvement of the estimate accuracy induced by the 2nd-level 
H-WT with respect to that obtained via the 1st-level one of the same 
order appears, at least for values of $N$ not too large, not to 
adequately refund the unavoidable increase of the computational 
complexity required by the application of the 2nd-level 
transformation. For example,
it is seen from Fig. \ref{FigErrorAiryF16} that a relative error of 
the order of $10^{-18}$, achieved through the 2nd-level H-WT with 
$N=11$ and $M=8$, would be 
reached via a 1st-order H-WT by letting $N=13$, but with a 
considerable saving of computational effort.
The above example clearly suggests the use of the 2nd-level H-WT only 
in those cases where the best accuracy attainable via the 1st-level 
H-WT 
turns out to be not adequate. This happens, for instance, 
when the integral is attempted to be evaluated beyond the 
asymptotic realm. 
To put into evidence this aspect,
Figure~\ref{FigErrorAiryF14To2} shows the same as in 
Fig.~\ref{FigErrorAiryF16}, but for {a decreasing sequence of values 
of $F$, namely}  
14 (a), 10 (b), 6 (c), and 2 (d). In particular, in 
Fig.~\ref{FigErrorAiryF14To2}(d), where
the Airy function argument is located at a distance $\simeq 1$ from 
the origin of the complex plane, the 1st-level H-WT provides a 
best error of the order of $10^{-4}$, achieved for $N=4$. Higher
accuracies are not allowed 
because the information gained at the first H stage turns out 
to be no longer sufficient to generate WT-resummable sequences.
The 2nd-level H-WT, on the other hand, provides a best error of 
the order of $10^{-10}$, which is attained for $(N,M)=(15,11)$.
Some intuitive insights about the resummation process associated to 
the H-WT could be grasped by noting that a lower bounded error is an 
intrinsic imprint of superasymptotic and hyperasymptotic resummations, 
whereas it is not generally featured by the application of the WT to 
alternating factorial divergent series\cite{wenigerCPR-89}. 
Accordingly, one should be inclined to think that such an error 
behavior could be ascribed to the presence of the ``regularization" 
step operated by H on the raw input data.
{
Speaking within a more general context, this should be somewhat related 
to the possible presence on nonanalytic, nonperturbative correction terms 
which cannot be grasped simply by resummation processes, but rather require
the use of ``generalized nonanalytic expansions" \cite{calicetiArXiv07}.
}
\\
In a second experiment concerning the Airy function, 
the asymptotic parameter $F$ is let running within the
interval $[2,4]$ and, for each value of $F$, an exhaustive search of 
the optimal values 
of the truncations $N$ and $(N,M)$, which minimize the 1st- and 
2nd-level relative errors, 
respectively, is operated. The results are shown in 
Fig.~\ref{FigOptimalErrorAiry}, where 
the optimal relative errors obtained via the 1st- (open circles) and 
the 2nd-level (dots)
H-WT are shown as functions of $F$. The  values  of the optimal 
truncation $N$ 
for the 1st-level H-WT are also reported, versus $F$, in 
Fig.~\ref{FigOptimalNAiry},
while those of $N$ and $M$, for the 2nd-level H-WT, in 
Fig.~\ref{FigOptimalNMAiry}(a) and (b), respectively. We will come 
back later on the above results.
\\
In concluding the present section, however, we want to provide a
table of explicit values obtained through the use of the 2nd-level
H-WT. We choose to evaluate the Airy function for $F=2$, for which
the optimal setting of the truncation parameters turns out to be
$(N,M)=(15,11)$. The preliminary step is the evaluation, through a simple WT, 
of the contribution, to the Airy integral, coming from the saddle 
$s_2$. The result is shown in Table \ref{table.1}. Furthermore, the subsequent action of the 
2nd-level H-WT on the saddle $s_1$ is shown in Table \ref{table.2}, where the complete estimates of the
Airy function provided are reported together with the corresponding values of the truncation $M \in [3,14]$, with $N=15$.
}

\subsection{Instanton integral}
\label{instanton}

The second numerical experiment concerns
the evaluation of the instanton integral
\begin{equation}
\mathcal{N}(k)=k^{1/2}\,\displaystyle\int_{-\infty}^{+\infty}\,
\exp[-k (s^2-1)^2]\,\mathrm{d}s,
\label{inst.1}
\end{equation}
with  $k>0$, 
{already considered in Ref. \cite{berryPW-93} as a simplified 
prototype for the modeling of instanton tunneling between symmetric double 
wells. It was shown in Ref. \cite{borghiPRE-08} that
the integral in Eq. (\ref{inst.1}) can be written 
as\cite{borghiPRE-08}
}
\begin{equation}
    \mathcal{N}(k)=2\,k^{1/2}\,
\mathrm{Re}\left\{\mathcal{I}(k)\right\},
\label{inst.2}
\end{equation}
where, by referring to Eq. (\ref{sd.1}), $g(s)=1$, $f(s)=(s^2-1)^2$, 
and where $\mathcal{C}$ is the steepest descent path connecting the 
points $-\mathrm{i}\infty$ and $+\infty$ via the lines 
$\mathrm{Im}\{s\}\le 0$ and $\mathrm{Re}\{s\}\ge 0$.
The complete saddle topology, as well as the expressions of the 
expanding coefficients associated to all saddles have been described 
in Ref.~\cite{borghiPRE-08}. In particular, there are three saddles, 
$s_1=-1$, $s_2=0$, and $s_3=1$, with $\mathcal{A}_{1}=\{2\}$,  
$\mathcal{A}_{2}=\{1,3\}$, and  $\mathcal{A}_{3}=\{2\}$.
{The saddles involved in the evaluation of $\mathcal{I}(k)$
are $s_2$ and $s_3$, but only the latter requires a H-WT treatment, since the associated
singulant is $F_{32}=1 >0$, and the corresponding asymptotic series turns out to
be nonalternating.}
Furthermore, $\gamma_{12}= 1$, $\gamma_{21}= \gamma_{23}= 0$, and  
$\gamma_{32}=1$, while we recall that the integral in Eq. (\ref{inst.1}) 
can be expressed in closed form via
\begin{equation}
    \mathcal{N}(k)= \displaystyle\frac{\pi\,\sqrt k}{2}\,\exp(-k/2)\,
\left[
I_{-1/4}\left(\displaystyle\frac k2\right) + 
I_{1/4}\left(\displaystyle\frac k2\right)
\right],
\label{inst.4}
\end{equation}
where $I_n(\cdot)$ denotes the $n$th-order modified Bessel function 
of the first kind. The first experiment concerns the evaluation of $\mathcal{N}(1/2)$ 
via the 1st- and the 2nd-level H-WT. In Fig.~\ref{FigInstantonek1ov2} it is seen how the 
2nd-level relative error is bounded, with a minimum value of the order of $10^{-3}$, achieved
for $(N, M)=(6,5)$. {On the opposite, the 1st-level H-WT turns 
out to be completely inadequate to provide a reasonably accurate
estimate of the function, due to the very low value of $k$.}
The searching for optimal values has also been carried out in the present case, but 
using, as the varying asymptotic parameter, $k \in [1/2,3]$, i.e., 
where the 1st-order H-WT displays the worst results in terms of 
accuracy, as shown in Fig.~2a of Ref.~\cite{borghiPRE-08}. The error 
values are shown in Fig.~\ref{FigOptimalErrorInstanton}, versus $k$, 
while the optimal settings of $N$ 
and of $(N,M)$ are plotted, against $k$, in 
Fig.~\ref{FigOptimalNInstanton} 
and~\ref{FigOptimalNMInstanton}, respectively.

{It is worth now comparing the results pertinent to the Airy and 
the $\mathcal{N}(k)$ functions. What we are going to show can seem
at first sight somewhat surprising, but gives a possible first hint toward the 
understanding of the H-WT mechanisms. For simplicity we shall refer to the 
1st-level transformation, but the results will apply also to 
higher-order levels. 
In Fig.~\ref{FigComparisonAiryInst}, 
the values of the relative error obtained for the Airy function 
(dots) and for the instanton function (solid curve) are plotted, versus 
$N$, when the values of the parameter $F$ and the parameter $k$ are 
numerically equal. In particular, figure (a) corresponds to $k=F=3$,
(b) to 7, (c) to 12, and (d) to 20. It is clearly seen that the 
behavior of the relative error follows basically the same law.
To give a possible explanation of this, 
in Fig.~\ref{FigCOmpAiryInst} a pictorial representation of the 
complete saddle network and the complex integration path involved in the evaluation of the 
Airy (a) and of the instanton (b) functions is plotted. In  both pictures, the black 
dot denotes the saddle for which the H-WT is required. 
Although the two saddle distributions are clearly different, they 
present some common features that, together with Eq. (\ref{Istage.complete}),
are enough to justify what happens in Fig.~\ref{FigComparisonAiryInst}.
Each of the ``black'' saddles is adjacent to a single saddle. For the 
Airy function $s_{1}$ is adjacent to $s_{2}$, while for the instanton 
function $s_{3}$ is adjacent to $s_{2}$. The values of the 
corresponding singulants are $F$ and 1, respectively. The use of the resurgence relation in 
Eq.~(\ref{resurgence.1}) now gives, for the two ``black'' saddles, 
\begin{equation}
\begin{array}{lcr}
T^{(1)}_r \propto \displaystyle\frac{(r-1)!}{F^r},
\end{array}
\label{airy1st}
\end{equation}
for the Airy function and 
\begin{equation}
\begin{array}{lcr}
T^{(3)}_r \propto \displaystyle\frac{(r-1)!}{k^r},
\end{array}
\label{inst1st}
\end{equation}
for the instanton function. From the above equation it is seen that 
the behavior of the expanding coefficients follows the same asymptotic
law as soon as $F=k$. At the same time, however, the above equality 
guarantees that also the asymptotic laws for the expanding coefficients 
corresponding to the adjacent saddles is identical. In fact, for the 
Airy function the saddle adjacent to $s_{2}$ is $s_{1}$ itself, with 
a singulant value of $-F$. As far as the instanton function is 
concerned, the saddles adjacent to $s_{2}$ are $s_{1}$ and $s_{3}$, 
but for both of them the singulant values are -1. Accordingly, the use
of Eq.~(\ref{resurgence.1}), together with the condition $F=k$, 
provides again an equivalence between the asymptotic laws of $T^{(2)}_{r}$ 
for the Airy function and $T^{(2)}_{r}$ for the instanton function. Finally, 
on using Eq. (\ref{Istage.complete}) it is not difficult to convince that 
the retrieving process is the same for the two functions at the 
1st-level\cite{endnote30}. Leaving a deeper understanding of this phenomenon 
to future investigations, it is here worthwhile to point out how an immediate
consequence of the above  described ``topological equivalence" could be the 
restriction of the study of the H-WT retrieving performances to a few classes of
prototype test cases. 
}

\subsection{Swallowtail diffraction catastrophe}
\label{sw}

As a last numerical experiment we consider asymptotic evaluations 
of the so-called swallowtail diffraction 
catastrophe \cite{berryPIO-80,connorJPA-84,nye,nyePRSA-07},
which is defined through the following integral:
\begin{equation}
S(x,y,z)=
\displaystyle\int_{\mathcal{C}}\,
\exp\left[\mathrm{i}\left(
\displaystyle\frac{s^5}5+x\,\displaystyle\frac{s^3}3+y\,\displaystyle\frac{s^2}2+zs\right)\right]\,
\mathrm{d}s,
\label{sw.1}
\end{equation}
which is of the form given in Eq. (\ref{sd.1}) with $g(s)=1$, 
$f(s)=-\mathrm{i} (s^5/5+x s^3/3+y s^2/2+zs)$, and $k=1$.
The integration path $\mathcal{C}$ can be thought of as
the union of steepest descent paths approaching, for $|s|\gg 1$, 
the directions $\varphi=(2n+1/2)\pi/5$, with $n=0,1,...,4$.
Although a systematic treatment of the swallowtail asymptotics, along the 
general classical rules recalled in Sec. \ref{notations},
can be derived by paralleling the analysis 
carried out, for the Pearcey function, in Ref.~\cite{berryPRSA-91},
up to our knowledge it is not present in the current literature. 
As shown in appendix  \ref{swT}, all expanding coefficients $T^{(n)}_r$  
are given by
\begin{equation}
\begin{array}{l}
T^{(n)}_r=
\displaystyle\frac
{(5\mathrm{i})^{r+1/2}\,(r-1/2)!}
{(10s^{3}_{n}+5s_{n}x+5y/2)^{5r/3+1/2}}\,
B^{(r+1/2)}_{2r}(\alpha,\beta),
\end{array}
\label{generatingC.2}
\end{equation}
where
\begin{equation}
\begin{array}{l}
\alpha=\displaystyle\frac{5s_{n}}{(10s^{3}_{n}+5s_{n}x+5y/2)^{1/3}},\\
\\
\beta=
\displaystyle\frac {10 s^{2}_{n} + 
{5x}/{3}}{(10s^{3}_{n}+5s_{n}x+5y/2)^{2/3}},
\end{array}
\label{sw.8.1.1.1.1.1.1.1.1}
\end{equation}
and where the polynomials $B^{(\lambda)}_{n}(u,v)$ are defined
via the  generating function formula
\begin{equation}
\begin{array}{l}
\displaystyle\sum_{n=0}^\infty\,
{t^n}\,B^{(\lambda)}_n(u,v)=
\displaystyle\frac 1{(t^3+u t^2+v t+1)^\lambda}.
\end{array}
\label{generatingC.3}
\end{equation}
It is not difficult to prove that
Eq. (\ref{generatingC.3}) allows the numerical evaluation of the 
polynomials $B^{(\lambda)}_n(u,v)$ to be efficiently performed 
via the use of the following recurrence rule, whose derivation
is outlined in Appendix \ref{recurrenceRule}:
\begin{equation}
\begin{array}{l}
nB^{(\lambda)}_n=-(n-3+3\lambda)\,B^{(\lambda)}_{n-3}-u\,(n-2+2\lambda)\,\,B^{(\lambda)}_{n-2}\\

\\-v\,(n-1+\lambda)\,B^{(\lambda)}_{n-1},
\end{array}
\label{generatingC.4}
\end{equation}
with the triggering values $B^{(\lambda)}_0(u,v)=1$, 
$B^{(\lambda)}_1(u,v)=-\lambda v$, and 
$B^{(\lambda)}_2(u,v)=-u\lambda+v^2\lambda(\lambda+1)/2$. 

The numerical experiments we are going to illustrate concern 
asymptotic evaluations of $S(x,y,z)$ at points belonging to the 
corresponding Stokes set,
{which has been estensively studied in Ref. \cite{berryNL-90} 
(see, in particular,
Fig. 3 of this reference)}.
Accordingly, the triplets $(x,y,z)$ have been chosen following the 
prescriptions given in Ref. \cite{berryNL-90}, in order to 
investigate points at
the intersection between the Stokes surface and the plane 
$x=0$, along the branch corresponding to
$y>0$ and $z>0$. This leads to triplets of the form 
$(x,y,z)=(0, \kappa^{3/2}, \kappa^2\times\,0.23012\ldots)$, with 
$\kappa$ being a positive parameter\cite{berryNL-90}. We start by 
considering the case $\kappa=2$. 
The saddle topology is  constituted by four saddles, which are listed 
below 
together with the corresponding list of adjacent ones:
\begin{equation}
\begin{array}{lcl}
s_1=0.8062\ldots- \mathrm{i}\,1.2357\ldots , & & 
\mathcal{A}_1=\{4\},\\
&&\\
s_2=s^*_1, & & \mathcal{A}_2=\{4\},\\
&&\\
s_3=-1.2828\ldots, & & \mathcal{A}_3=\{4\},\\
&&\\
s_4=-0.3296\ldots, & & \mathcal{A}_4=\{1,2,3\},
\end{array}
\label{sw.1.1}
\end{equation}
with the orientation matrix being
\begin{equation}
\{\gamma_{nm}\}=
\left[
\begin{array}{cccc}
\cdot & \cdot & \cdot & 1  \\
\cdot & \cdot & \cdot & 0  \\
\cdot & \cdot & \cdot & 0  \\
0 & 1 & 1 & \cdot  
\end{array}
\right],
\label{sw.1.2}
\end{equation}
{where the values of $\gamma_{nm}$ which are not relevant for the 
present experiment have been replaced by dots.}
{Three of the four saddles, namely $s_{2}$, $s_{3}$ and 
$s_{4}$, do contribute to the 
integral. In particular,}
the integration path consists in the union of three steepest-descent 
arcs, the first
connecting the points $\infty\,\exp(\mathrm{i}9\pi/10)$ and 
$\infty\,\exp(\mathrm{i}13\pi/10)$ passing through $s_3$, the second  connecting the point 
$\infty\,\exp(\mathrm{i}13\pi/10)$ to the saddle $s_2$, passing through $s_4$, and the third connecting the saddle $s_2$ to the point $\infty\,\exp(\mathrm{i}\pi/10)$.  Accordingly, the Stokes phenomenon occurs via the so-called \emph{saddle connection} 
between saddles $s_4$ and $s_2$\cite{berryNL-90}. 
{A pictorial representation of the topology above described is given, for the reader convenience, in Fig. \ref{FigSwPaths}.}

{Before showing the numerical results about the performances of 
the 1st- and the 2nd-level H-WT, it is worth giving some details 
about the way the saddle topology of the swallowtail integral influences 
the retrieving capabilities of the WT. We refer, in particular, to 
the contribution, to the swallowtail integral, of $s_{3}$ and 
$s_{4}$. As far as the former is concerned, since there is only one 
adjacent saddle, $s_{4}$,
due to the fact that the corresponding singulant 
$F_{34}=\mathrm{i}\,0.602168\ldots$
is purely imaginary, it turns out that $T^{(3)}_{r} \propto
(-\mathrm{i})^{r}\,(r-1)!/|F_{3,4}|^{r}$, thus allowing the WT to 
operate the resummation. Similar considerations can be 
done for the contributive saddle $s_{2}$.
The situation is somewhat different for the saddle $s_{4}$, which is 
connected to $s_{2}$. In fact, 
from the above described topology, it turns out that the adjacent 
saddle $s_{3}$
is dominant, i.e., presents the minimum value of $|F_{4,m}|$, with $m 
\in 
\mathcal{A}_{4}$. Accordingly, one should conclude also for 
$T^{(4)}_{r}$
an asymptotic ``factorial divided by power'' law, similar to that 
corresponding to  $T^{(3)}_{r}$ and, due to the fact that
$F_{4,3}=-\mathrm{i}\,0.602168\ldots$, one should expect the WT to 
be able in resumming the corresponding asymptotic series. 
But this, on the contrary, does not happen, because of the 
presence of the other two, nondominant, saddles $s_{1}$ and $s_{2}$, 
symmetrically placed in the complex singulant space.
This can be explained by expliciting Eq.~(\ref{resurgence.1}) as
\begin{equation}
T^{(4)}_r\approx \displaystyle\frac {(r-1)!}{2\pi\mathrm{i}}\,
\left[
-\displaystyle\frac{T^{(3)}_{0}}{F^{r}_{43}}+
\left(
\displaystyle\frac{T^{(1)}_{0}}{F^{r}_{41}}-
\displaystyle\frac{T^{(2)}_{0}}{F^{r}_{42}}
\right)
\right],
\label{resurgence.1Bis}
\end{equation}
where $T^{(2)}_{0}=\mathrm{i}[T^{(1)}_{0}]^{*}$.
This example shows how the divergent asymptotic series
generated on Stokes sets do not necessarily display a strictly 
nonalternating sign pattern as, for example, happened for the Airy 
function, but rather how the asymptotic behavior of their single 
terms 
can display more complex patterns, depending on 
the whole saddle topology.
Figure~\ref{FigSWk2} shows the relative errors obtained through 
the 1st- and the 2nd-level H-WTs in the case of swallowtail 
evaluation 
across the Stokes set defined above, when $\kappa=2$.
As for Figs.~\ref{FigErrorAiryF16},~\ref{FigErrorAiryF14To2}, and~
\ref{FigInstantonek1ov2}, the errors are plotted versus the 
parameters $N$ and $M$. In this case 
their optimal values turn out to be
$N=4$ (for the 1st-level) and $(N,M)=(7,6)$ (for the 2nd-level),
with corresponding error values of $2\cdot10^{-3}$ and 
$5\cdot10^{-5}$, 
respectively, evaluated with respect to the ``exact" value of 
$S(0,2.8284\ldots,0.9205\ldots)$, obtained via the method recently 
proposed in Ref. \cite{borghiJOSAA-08bis}.
Finally, an experiment about optimal resummation of swallowtail 
functions has been carried out, by using $\kappa \in [2,4]$ as the 
parameter
representative of the ``asymptoticity" features.
The error values are shown, versus $\kappa$, in 
Fig.~\ref{FigOptimalErrorsw}, 
while the optimal settings of $N$ and of $(N,M)$ are 
plotted, against $\kappa$, in Fig.~\ref{FigOptimalNsw}
and~\ref{FigOptimalNMsw}, respectively.
}

\section{Conclusions}
\label{conclusions}

{The H-WT was introduced in Ref.~\cite{borghiPRE-08} as a 
powerful and easily implementable refinement of H
aimed at allowing the WT to successfully decode 
those divergent asymptotic series generated 
through the application of the steepest descent method 
to saddle point integrals evaluated across Stokes sets,
for which their single terms do not display a strictly alternating 
sign 
pattern.
}
The scheme proposed in \cite{borghiPRE-08} employed the WT only on 
the asymptotic series generated by the first-stage 
hyperasymptotic treatment of the corresponding diverging remainder. 
In the present sequel we reported on the possibility of
combining H and the WT to higher orders in H. In particular, 
the full development of the 2nd-level H-WT has been detailed 
within the classical framework of H.
The results obtained from the application of the 2nd-level H-WT, 
also in comparison to those obtained via the 1st-level one, on the 
different types of saddle-point integrals considered, showed how 
the increase of complexity and computational effort required
by the new transformation be adequately refunded, in terms of 
accuracy of the estimate, particularly when the integrals are 
evaluated for 
values of their parameters which are beyond the asymptotic regime, 
whereas H turns out to be inapplicable and the 1st-order H-WT 
unavoidably lacks 
precision.
At the same time, however, it should be noted how, for ``ordinary"
asymptotic evaluations, at least in the cases considered in the 
present
work, the performances of the 1st- and the 2nd-level H-WTs seem to be 
comparable in terms of the estimate accuracy, against a considerably 
difference in the computational efforts required by the two 
transformations.

{Although the H-WT has been developed, here and in 
Ref.~\cite{borghiPRE-08},
with reference to the evaluation of saddle-point integrals of the 
type in
Eq.~(\ref{sd.1}), we believe it could be useful also in dealing with
problems of different nature like, for instance, the 
hyperasymptotic treatment of a wide class of linear and nonlinear 
ordinary differential equations, which has been recently 
considered~\cite{berryJPA-99,daalhuisPRSA-05,daalhuisPRSA-05b}.
The semi-analytical algorithms proposed 
for the numerical evaluation of the 2nd-level hyperterminants would 
reveal useful in this perspective.}

Before concluding, it is worth pointing out, 
as an important open problem, the need for an 
\emph{a priori} algorithm for estimating the values of 
$N$ and $M$ that lead to optimal results. Especially
in cases where it is not convenient (or possible) to 
evaluate the original function, such an algorithm would certainly 
be of great help to a typical user.
Unfortunately, differently from H, for which the optimal settings of 
the hyperseries truncations are directly extracted from the singulant 
values\cite{berryPRSA-90,berryPRSA-91}, at present it does not seem 
possible to provide similar information for the H-WT. 
The difficulties in giving practical guidelines for the choice of $N$ 
and $(N,M)$ can also be appreciated from the results presented in 
Sec. \ref{numericalExperiments} 
and especially from Figs. \ref{FigOptimalNAiry}, 
\ref{FigOptimalNMAiry}, \ref{FigOptimalNInstanton}, 
\ref{FigOptimalNMInstanton}, \ref{FigOptimalNsw}, and 
\ref{FigOptimalNMsw}, where it seems quite difficult to obtain general 
rules for the optimal settings of them. 
{A possible hint, grasped from a quantitative comparison between the
results obtained for the Airy and the instanton functions, seems to be given by 
the strong connection between the H-WT retrieving performances and
the saddle topology associated to the integral under consideration.
What we found is that different saddle networks can share a sort of 
``topological equivalence" property, which is related to the set of 
the saddles adjacent to that under consideration and to the  
values of the relevant singulants. If two networks turn out to be 
equivalent at a certain hyperasymptotic level, this would result in the same computational
effort, in terms of relative error, as far as the corresponding  H-WT retrieved estimates
are concerned. This, in particular, would imply that the study of the H-WT retrieving performances 
could be, in principle, carried out only on a restricted class of prototype functions.
}

\acknowledgments

I would like to thank all anonymous reviewers for their constructive 
criticisms
and suggestions. I am also grateful to Turi Maria Spinozzi for his 
invaluable help
during the preparation of the present work.

\appendix

\section{Second-stage hyperasymptotics}
\label{IIstage}

The starting point is again the integral representation of the 
remainder $R^{(n)}(k,N)$ given in Eq. (\ref{remainder.2}),
where we let
\begin{equation}
T^{(m)}\left(\displaystyle\frac v{F_{nm}}\right)=
\displaystyle\sum_{r=0}^{M-1}\,\displaystyle\frac{v^{-r}}{F^{-r}_{nm}}\,T^{(m)}_r+
\displaystyle\sum_{r=M}^{\infty}\,\displaystyle\frac{v^{-r}}{F^{-r}_{nm}}\,T^{(m)}_r,
\label{IIstage.1}
\end{equation}
with $M$ being a \emph{new} truncation order. Substitution of 
Eq. (\ref{IIstage.1}) into Eq. (\ref{remainder.2})
gives 
\begin{equation}
\begin{array}{lcl}
R^{(n)}(k,N) =  
\displaystyle\frac {(-1)^N}{2\pi\mathrm{i}}\,
\displaystyle\sum_{m \in \mathcal{A}_n}\,(-1)^{\gamma_{nm}}\,\\
\\
\times
\displaystyle\sum_{r=0}^{M-1}\,
(-1)^r\,k^{-r}\,T^{(m)}_r\,K^{(1)}_{N-r}(-kF_{nm}),\\
\\+
\displaystyle\frac {1}{2\pi\mathrm{i}}\,
\displaystyle\sum_{m \in \mathcal{A}_n}\,(-1)^{\gamma_{nm}}\\
\\
\times
\displaystyle\sum_{r=M}^{\infty}\,\displaystyle\frac{(-1)^{N-r}}{(-kF_{nm})^{N-r}}\,k^{-r}\,T^{(m)}_r\,

\displaystyle\int_0^\infty\,
\mathrm{d}u\,
\exp(-u)\,
\displaystyle\frac{u^{N-r-1}}{1-\displaystyle\frac{u}{kF_{nm}}}.
\end{array}
\label{IIstage.2}
\end{equation}
On using resurgence, i.e., Eq. (\ref{resurgence.1}), on $T^{(m)}_r$, 
namely
\begin{equation}
\begin{array}{lcl}
T^{(m)}_r= \displaystyle\frac {1}{2\pi\mathrm{i}}\,
\displaystyle\sum_{h \in \mathcal{A}_m}\,(-1)^{\gamma_{mh}}
\displaystyle\sum_{l=0}^\infty\,
\displaystyle\frac{(r-l-1)!}{F^{r-l}_{mh}}\,T^{(h)}_l,
\end{array}
\label{IIstage.3}
\end{equation}
and 
\begin{equation}
\begin{array}{lcl}
(r-l-1)!= \displaystyle\int_0^\infty\,
\mathrm{d}v\,
\exp(-v)\,v^{r-l-1},
\end{array}
\label{IIstage.3.0.1}
\end{equation}
after straightforward algebra
the second term in Eq. (\ref{IIstage.2}) becomes
\begin{equation}
\begin{array}{lcl}
\left(\displaystyle\frac {1}{2\pi\mathrm{i}}\right)^2\,
\displaystyle\sum_{m \in \mathcal{A}_n}
\displaystyle\sum_{h \in 
\mathcal{A}_m}\,(-1)^{\gamma_{nm}+\gamma_{mh}}\,
\displaystyle\sum_{l=0}^\infty\,k^{-l}T^{(h)}_l\\
&&\\
\times 
\displaystyle\sum_{r=M}^\infty\,
\displaystyle\frac 1{(kF_{nm})^{N-r}}\,
\displaystyle\frac 1{(kF_{mh})^{r-l}}\,\\
\\
\times
\displaystyle\int_0^\infty\,
\displaystyle\int_0^\infty\,
\displaystyle\frac{\mathrm{d}u}{u}\,
\displaystyle\frac{\mathrm{d}v}{v}\,
\exp(-u-v)\,\displaystyle\frac{u^{N-r}\,v^{r-l}}{1-\displaystyle\frac 
u{kF_{nm}}},
\end{array}
\label{IIstage.3.1}
\end{equation}
or, by interchanging the integral with the $r$-sum,
\begin{equation}
\begin{array}{lcl}
\left(\displaystyle\frac {1}{2\pi\mathrm{i}}\right)^2\,
\displaystyle\sum_{m \in \mathcal{A}_n}
\displaystyle\sum_{h \in 
\mathcal{A}_m}\,(-1)^{\gamma_{nm}+\gamma_{mh}}\,
\displaystyle\sum_{l=0}^\infty\,k^{-l}T^{(h)}_l\,
\displaystyle\frac{(kF_{mh})^{l}}{(kF_{nm})^{N}}\\
&&\\
\times 
\displaystyle\int_0^\infty\,
\displaystyle\int_0^\infty\,
\displaystyle\frac{\mathrm{d}u}{u}\,
\displaystyle\frac{\mathrm{d}v}{v}\,
\exp(-u-v)\,\displaystyle\frac{u^{N}\,v^{-l}}{1-\displaystyle\frac 
u{kF_{nm}}}\,\\
\\
\times
\displaystyle\sum_{r=M}^\infty\,
\left(\displaystyle\frac 
vu\,\displaystyle\frac{F_{nm}}{F_{mh}}\right)^{r}
=\\
&&\\
\left(\displaystyle\frac {1}{2\pi\mathrm{i}}\right)^2\,
\displaystyle\sum_{m \in \mathcal{A}_n}
\displaystyle\sum_{h \in 
\mathcal{A}_m}\,(-1)^{\gamma_{nm}+\gamma_{mh}}\,
\displaystyle\sum_{l=0}^\infty\,k^{-l}T^{(h)}_l\,
\displaystyle\frac{(kF_{mh})^{l}}{(kF_{nm})^{N}}\\
&&\\
\times 
\displaystyle\int_0^\infty\,
\displaystyle\int_0^\infty\,
\displaystyle\frac{\mathrm{d}u}{u}\,
\displaystyle\frac{\mathrm{d}v}{v}\,
\exp(-u-v)\,\displaystyle\frac{u^{N}\,v^{-l}}{1-\displaystyle\frac 
u{kF_{nm}}}\,
\displaystyle\frac{\left(\displaystyle\frac 
vu\,\displaystyle\frac{F_{nm}}{F_{mh}}\right)^{M}}{1-\left(\displaystyle\frac 
vu\,\displaystyle\frac{F_{nm}}{F_{mh}}\right)}.
\end{array}
\label{IIstage.3.2}
\end{equation}
Eventually, on making the substitution $-v F_{nm}/F_{mh} \to v$, 
after some algebra the quantity in Eq. (\ref{IIstage.3.2})
can be written as
\begin{equation}
\begin{array}{lcl}
\displaystyle\frac{(-1)^{N+M}}{(2\pi\mathrm{i})^2}\,
\displaystyle\sum_{m \in \mathcal{A}_n}
\displaystyle\sum_{h \in 
\mathcal{A}_m}\,(-1)^{\gamma_{nm}+\gamma_{mh}}\,\\
\\
\times
\displaystyle\sum_{l=0}^\infty\,k^{-l}T^{(h)}_l
K^{(2)}_{M-l,N-M}\left(-kF_{nm};-\displaystyle\frac{F_{mh}}{F_{nm}}\right),

\end{array}
\label{IIstage.4}
\end{equation}
where the 2nd-level hyperterminant $K^{(2)}_{n,m}(\beta;\gamma)$, 
defined by
\begin{equation}
\begin{array}{lcl}
K^{(2)}_{n,m}(\beta;\gamma) = 
\displaystyle\frac 1{\beta^{n+m}}\,\\
\\
\times
\displaystyle\int_0^\infty\,
\displaystyle\int_0^\infty\,
\mathrm{d}u\,\mathrm{d}v\,
\exp(-u-\gamma\,v)\,
\displaystyle\frac{u^{m-1} v^{n-1}}{\left(1+\displaystyle\frac 
u\beta\right)\,\left(1+\displaystyle\frac v u \right)},
\end{array}
\label{IIstage.6}
\end{equation}
has been introduced. Accordingly, the complete expression of the 
2nd-level hyperasymptotic expansion is
\begin{equation}
\begin{array}{lcl}
T^{(n)}(k)=\displaystyle\sum_{r=0}^{N-1}\,k^{-r}\,T^{(n)}_r\\
\\+
\displaystyle\frac {(-1)^N}{2\pi\mathrm{i}}\,
\displaystyle\sum_{m \in \mathcal{A}_n}\,(-1)^{\gamma_{nm}}\\
\\
\,\,\,\,\,\,\,\,\,\,\,\,\,\,\,\,\,\,\,\,\,\times
\displaystyle\sum_{r=0}^{M-1}\,
(-1)^r\,k^{-r}\,T^{(m)}_r\,K^{(1)}_{N-r}(-kF_{nm})\\
\\
+
\displaystyle\frac{(-1)^{N+M}}{(2\pi\mathrm{i})^2}\,
\displaystyle\sum_{m \in \mathcal{A}_n}
\displaystyle\sum_{h \in 
\mathcal{A}_m}\,(-1)^{\gamma_{nm}+\gamma_{mh}}\,\\
\\
\,\,\,\,\,\,\,\,\,\,\,\,\,\,\,\,\,\,\,\,\,\,\,\,\,\,\,\,\,\,\times
\displaystyle\sum_{r=0}^\infty\,k^{-r}T^{(h)}_r\,
K^{(2)}_{M-r,N-M}\left(-kF_{nm};-\displaystyle\frac{F_{mh}}{F_{nm}}\right),

\end{array}
\label{IIstage.complete.bis}
\end{equation}
which coincides with Eq. (\ref{IIstage.complete}).

\section{Derivation of the extra term for $\beta <0$}
\label{appC}

{
We first recast the double integral in Eq. (\ref{terminant2}) as 
follows:
\begin{equation}
\begin{array}{l}
K^{(2)}_{n,m}(\beta;\gamma) = \displaystyle\frac 1{\beta^{n+m-1}}\,
\displaystyle\int_0^\infty\,
\mathrm{d}v\,\exp(-\gamma v)\,v^{n-1}\,\\
\\
\times
\displaystyle\int_0^\infty\,
\mathrm{d}u\,\exp(-u)\,
\displaystyle\frac{u^{m}}{(u+\beta)(u+v)},
\end{array}
\label{betaNegativo.1}
\end{equation}
which, if $\beta$ were not negative, would take the form
\begin{equation}
\begin{array}{l}
K^{(2)}_{n,m}(\beta;\gamma) = \displaystyle\frac {m!}{\beta^{n+m-1}}\,
\displaystyle\int_0^\infty\,
\mathrm{d}v\,\exp(-\gamma v)\,v^{n-1}\\
\\
\times
\left[\displaystyle\frac{\exp(v)E_{m+1}(v)-\exp(\beta)E_{m+1}(\beta)}{\beta-v}\right],

\end{array}
\label{betaNegativo.2}
\end{equation}
where the integrand turns out to be continuous even for real, 
positive values of $\beta$, since
\begin{equation}
\begin{array}{l}
    \displaystyle\lim_{v\to\beta}\,
\displaystyle\frac{\exp(v)E_{m+1}(v)-\exp(\beta)E_{m+1}(\beta)}{\beta-v}=\\
\\
=
\displaystyle\frac{1-\exp(\beta)\beta^{m}(m+\beta)\Gamma(-m,\beta)}{\beta}.
\end{array}
\label{betaNegativo.2.1}
\end{equation}

If $\beta <0$, Eq.~(\ref{betaNegativo.1}) must be written as
\begin{equation}
    \begin{array}{l}
    K^{(2)}_{n,m}(\beta;\gamma) = \displaystyle\frac 
1{\beta^{n+m-1}}\,
    \displaystyle\int_0^\infty\,
    \mathrm{d}v\,\exp(-\gamma v)\,v^{n-1}\,\\
    \\
    \times
    \mathcal{P}\displaystyle\int_0^\infty\,
    \mathrm{d}u\,\exp(-u)\,
    \displaystyle\frac{u^{m}}{(u+\beta)(u+v)},
    \end{array}
\label{betaNegativo.4}
\end{equation}
where $\mathcal{P}\displaystyle\int_{0}^{\infty}\,\ldots$ denotes
the principal part operator. On evaluating the integral in 
Eq.~(\ref{betaNegativo.4}) we have
\begin{equation}
    \begin{array}{l}
    \mathcal{P}\displaystyle\int_0^\infty\,
    \mathrm{d}u\,\exp(-u)\,
    \displaystyle\frac{u^{m}}{(u+\beta)(u+v)}=\\
    \\
    =
    {m!}\,
\left[\displaystyle\frac{\exp(v)E_{m+1}(v)-\exp(\beta)E_{m+1}(\beta)}
{\beta-v}\right]\\
    \\-
\mathrm{i}\pi\exp(\beta)\,\displaystyle\frac{(-\beta)^{m}}{\beta-v}.
    \end{array}
\label{betaNegativo.5}
\end{equation}
Finally, the direct substitution of Eq.~(\ref{betaNegativo.5}) into 
Eq.~(\ref{betaNegativo.4})  leads to the closed-form of the extra 
term as
\begin{equation}
    \begin{array}{l}
	\displaystyle\frac 
{\mathrm{i}\pi\exp(\beta)(-\beta)^{n}}{\beta^{n+m-1}}\,
	    \displaystyle\int_0^\infty\,
	    \mathrm{d}v\,\exp(-\gamma v)\,
	    \displaystyle\frac{v^{n-1}}{v-\beta}=\\
	    \\
=\mathrm{i}\pi\,(-1)^{n+m-1}(n-1)!\displaystyle\frac{\exp[\beta(1-\gamma)]}
{(-\beta)^{n-1}}
	    \,E_{n}(-\beta\gamma).
	\end{array}
\label{betaNegativo.6.1}
\end{equation}
}

\section{Asymptotic coefficients $T^{(n)}_r$ for 
the swallowtail function}
\label{swT}

The swallowtail function in Eq. (\ref{sd.1}) corresponds to 
let $g=1$, $k=1$, and 
$f(s)=-\mathrm{i}(s^{5}/5+xs^{3}/3+ys^{2}/2+zs)$
in Eq. (\ref{sw.1}).
The (four) saddle points $\{s_n\}$ are solutions of the algebraic 
equation
\begin{equation}
s^4+x\,s^{2}+y\,s+z=0,
\label{sw.1.1.1.1.1}
\end{equation}
which can be exactly solved by using, for instance, Cardano's 
formula. 
The evaluation of the expanding coefficients $T^{(n)}_r$
requires to evaluate the integral in Eq.~(\ref{sdReview.5.0}).
The first step is to expand $f(s)-f_n$ around $s_n$, i.e.,
\begin{equation}
\begin{array}{lll}
f(s)-f_n= \\
\\=
-\mathrm{i}\left[\displaystyle\frac{s^{5}-s_n^{5}}5+
x\,\displaystyle\frac{s^{3}-s_n^{3}}3+
y\,\displaystyle\frac{s^{2}-s_n^{2}}2+
z\,({s-s_n})\right]=\\
\\
=\displaystyle\frac{(s-s_n)^2}{5\mathrm{i}}\,P(s),
\end{array}
\label{sw.3}
\end{equation}
where
\begin{equation}
P(s)=
s^{3}+
2s_{n}s^{2}+
\left(3s^{2}_{n}+\displaystyle\frac 53x\right)\,s+
\left(4s^{3}_{n}+\displaystyle\frac {10 s_{n}}3x+\displaystyle\frac 
52y\right).
\label{sw.3.1}
\end{equation}
Substitution into Eq. (\ref{sdReview.5.0}) gives
\begin{equation}
\begin{array}{l}
T^{(n)}_r  = \\
\\=
\displaystyle\frac{(r-1/2)!\,\,\,(5\,\mathrm{i})^{r+1/2}}{2\pi\mathrm{i}}\,
\displaystyle\oint_{n}\,\displaystyle\frac{\mathrm{d}s}
{(s-s_n)^{2r+1}\,[P(s)]^{r+1/2}}=\\
\\
=\displaystyle\frac{(r-1/2)!\,\,\,(5\,\mathrm{i})^{r+1/2}\,}{(2r)!}\,
\left\{
\displaystyle\frac{\mathrm{d}^{2r}}{\mathrm{d} s^{2r}}\,
\displaystyle\frac 1{[P(s)]^{r+1/2}}
\right\}_{s=s_n}.
\end{array}
\label{sw.5}
\end{equation}
Equation~(\ref{sw.5}) can be further simplified by letting 
$u=s-s_n$ and then by changing $u$ into 
$t=u/(10s^{3}_{n}+5s_{n}x+5y/2)^{1/3}$,
which yields
\begin{equation}
\begin{array}{l}
T^{(n)}_r=
\displaystyle\frac
{(5\mathrm{i})^{r+1/2}\,\,\,(r-1/2)!}
{(10s^{3}_{n}+5s_{n}x+5y/2)^{5r/3+1/2}}\\
\\
\times
\displaystyle\frac{1}{ (2r)!}\,
\left[\displaystyle\frac{\mathrm{d}^{2r}}{\mathrm{d} t^{2r}}\,
\displaystyle\frac{1}{(t^3+\alpha t^2+\beta t +1)^{r+1/2}},
\right]_{t=0},
\end{array}
\label{generatingC.1.5.1}
\end{equation}
where
\begin{equation}
\begin{array}{l}
\alpha=\displaystyle\frac{5s_{n}}{(10s^{3}_{n}+5s_{n}x+5y/2)^{1/3}},\\
\\
\beta=
\displaystyle\frac {10 s^{2}_{n} + 
{5x}/{3}}{(10s^{3}_{n}+5s_{n}x+5y/2)^{2/3}}.
\end{array}
\label{sw.8.1}
\end{equation}
Note that Eq.~(\ref{generatingC.1.5.1}) can also be given the 
alternative following form:
\begin{equation}
\begin{array}{l}
T^{(n)}_r=
\displaystyle\frac
{(5\mathrm{i})^{r+1/2}\,(r-1/2)!}
{(10s^{3}_{n}+5s_{n}x+5y/2)^{5r/3+1/2}}\,
B^{(r+1/2)}_{2r}(\alpha,\beta),
\end{array}
\label{generatingC.2.1}
\end{equation}
where the function $B^{(n)}_{\lambda}(u,v)$ is defined through
\begin{equation}
\begin{array}{l}
\displaystyle\sum_{n=0}^\infty\,
{t^n}\,B^{(\lambda)}_n(u,v)=
\displaystyle\frac 1{(t^3+u t^2+v t+1)^\lambda},
\end{array}
\label{generatingC.3.1}
\end{equation}
which coincides with Eq. (\ref{generatingC.3}).

\section{Derivation of the recurrence rule in Eq. (\ref{generatingC.4})}
\label{recurrenceRule}

The starting point is the definition of $B^{(\lambda)}_n(u,v)$
through the generating function in Eq. (\ref{generatingC.3.1}) which,
once derived with respect to $t$, gives
\begin{equation}
\begin{array}{l}
\displaystyle\sum_{n=0}^\infty\,
t^{n}\,n\,B^{(\lambda)}_n(u,v)=-\lambda
\displaystyle\frac {3t^3+2ut^2+vt}{(t^3+u t^2+v t+1)^{\lambda+1}},
\end{array}
\label{recurrenceRule.1}
\end{equation}
and, by taking Eq. (\ref{generatingC.3.1}) into account once again, 
leads to
\begin{equation}
\begin{array}{l}
(3\lambda t^3+2\lambda ut^2+\lambda vt)\,
\displaystyle\sum_{n=0}^\infty\,
t^{n}\,B^{(\lambda)}_n(u,v)\\
\\+
(t^3+u t^2+v t+1)\,\displaystyle\sum_{n=0}^\infty\,
t^{n}\,n\,B^{(\lambda)}_n(u,v)=0.
\end{array}
\label{recurrenceRule.2}
\end{equation}
By operating the products term by term, by rearranging 
the series indices, after long but straightforward algebra
Eq. (\ref{generatingC.4}) follows.

\newpage\

\newpage\

\section*{List of tables}

\begin{table}[!ht]
    \begin{tabular}{c|c}
    \hline
    WT order &  Contribution of saddle $s_2$ \\ \hline  \hline
2 & \underline{0.023}15166515+\underline{0.0}4009986036 $\mathrm{i}$\\
3 & \underline{0.0230}3913417+\underline{0.0399}0495099 $\mathrm{i}$\\
4 & \underline{0.02305}968041+\underline{0.0399}4053805 $\mathrm{i}$\\
5 & \underline{0.0230583}9073+\underline{0.039938}30429 $\mathrm{i}$\\
6 & \underline{0.023058}29942+\underline{0.0399381}4613 $\mathrm{i}$\\
7 & \underline{0.0230583}0901+\underline{0.03993816}279 $\mathrm{i}$\\
8 & \underline{0.0230583106}8+\underline{0.039938165}66 $\mathrm{i}$\\
9 & \underline{0.0230583106}8+\underline{0.039938165}62 $\mathrm{i}$\\
10 & \underline{0.02305831064}+\underline{0.03993816557} $\mathrm{i}$\\
11 & \underline{0.02305831064}+\underline{0.03993816557} $\mathrm{i}$\\
   ... & ...  \\
   	\hline
    \hline
    \end{tabular}
\caption{
{Values, provided by the WT, of the contribution coming from the saddle $s_2$ in the evaluation of 
the Airy function across its Stokes set, when $F=2$.
}
}
\label{table.1}
\end{table}
\begin{table}[!ht]
    \begin{tabular}{c|c}
    \hline
    M &  H-WT estimate \\ \hline  \hline
3 & \underline{0.7}906105793-0.4032083434 $\mathrm{i}$ \\
4 & \underline{0.70}61079570-\underline{0.35}44207316 $\mathrm{i}$ \\
5 & \underline{0.701}8070334-\underline{0.351}9375922 $\mathrm{i}$ \\
6 & \underline{0.7015}955837-\underline{0.3518}155117 $\mathrm{i}$ \\   	
7 & \underline{0.70158}34126-\underline{0.351808}4847 $\mathrm{i}$ \\   	
8 & \underline{0.701582}6207-\underline{0.3518080}275 $\mathrm{i}$ \\	
9 & \underline{0.701582}5835-\underline{0.3518080}060 $\mathrm{i}$ \\	
10 & \underline{0.701582}5939-\underline{0.35180801}20 $\mathrm{i}$ \\
11 & \underline{0.70158260}51-\underline{0.351808018}5 $\mathrm{i}$ \\	
12 & \underline{0.7015826}244-\underline{0.3518080}296 $\mathrm{i}$ \\	
13 & \underline{0.7015}563466-\underline{0.3517}928581 $\mathrm{i}$ \\
14 & \underline{0.7015826}472-\underline{0.3518080}428 $\mathrm{i}$ \\
    \hline
    exact & 0.7015826047 - 0.3518080182 $\mathrm{i}$\\
    \hline
    \end{tabular}
\caption{
{Estimates, provided by the 2nd-level H-WT, of the Airy function
across its Stokes set, when $F=2$. Note that the truncation parameter $N$
has been fixed to 15 (corresponding, from Fig. \ref{FigOptimalNMAiry}, to the optimal setting).
}
}
\label{table.2}
\end{table}

\newpage\
\pagebreak\

\section*{List of figure captions}

\begin{figure}[!ht]
\caption{
{
Behavior of the relative error, obtained through the 1st-level 
H-WT (dots), versus the values of $N$, which are reported on 
the abscissa axis. For each value of $N$, the values of the 
relative error obtained via the 2nd-level H-WT, with $M \in [3,N-1]$, 
are also plotted and, for the sake of clarity, are joined with lines 
of different color, each of them corresponding to a different value 
of $N$, departing from the abscissa $N$ itself.
}
}
\label{FigErrorAiryF16}
\end{figure}
\begin{figure}[!ht]
\caption{
{
The same as in Fig.~\ref{FigErrorAiryF16}, but for 
$F=14$ (a), 10 (b), 6 (c), and 2 (d).
}
}
\label{FigErrorAiryF14To2}
\end{figure}
\begin{figure}[!ht]
\caption{
{
Behaviors, as functions of $F\in[2,4]$, of the relative errors for 
the Airy function obtained via the 1st- (open circles) and the 
2nd-level (dots) H-WT. Each point has been obtained by carrying out, 
for a given $F$, an exhaustive search for those values of the 
truncations $N$ and $(N,M)$ which minimize the corresponding 1st- and 
2nd-level relative errors, respectively.
}
}
\label{FigOptimalErrorAiry}
\end{figure}
\begin{figure}[!ht]
\caption{
{
Behavior, as a function of $F$, of the values of
$N$ corresponding to the optimal setting for the 
1st-level H-WT in the experiment of Fig. \ref{FigOptimalErrorAiry}.	
}
}
\label{FigOptimalNAiry}
\end{figure}
\begin{figure}[!ht]
\caption{
{
Behavior, as a function of $F$, of the values of
$N$ (a) and $M$ (b) corresponding to the optimal setting for the 
2nd-level H-WT in the experiment of Fig. \ref{FigOptimalErrorAiry}.	
}
}
\label{FigOptimalNMAiry}
\end{figure}
\begin{figure}[!ht]
\caption{
{
The same as in Fig.~\ref{FigErrorAiryF16}, but for the instanton 
integral $\mathcal{N}(1/2)$.
}
}
\label{FigInstantonek1ov2}
\end{figure}
\begin{figure}[!ht]
\caption{
{
Behaviors, as functions of $k\in[1/2,3]$, of the relative errors for 
the instanton integral $\mathcal{N}(k)$ obtained via the 1st- (open 
circles) and the 2nd-level (dots) H-WT. As for the results presented 
in Fig. \ref{FigOptimalErrorAiry}, each point has been obtained by 
carrying out, for a given $k$, an exhaustive search for those values 
of the truncations $N$ and $(N,M)$ which minimize the corresponding 
1st- and 2nd-level relative errors, respectively.
}
}
\label{FigOptimalErrorInstanton}
\end{figure}
\begin{figure}[!ht]
\caption{
{
Behavior, as a function of $k$, of the values of
$N$ corresponding to the optimal setting for the 1st-level H-WT in 
the experiment of Fig. \ref{FigOptimalErrorInstanton}.	
}
}
\label{FigOptimalNInstanton}
\end{figure}
\begin{figure}[!ht]
\caption{
{
Behavior, as a function of $k$, of the values of
$N$ (a) and $M$ (b) corresponding to the optimal setting for the 
2nd-level H-WT in the experiment of Fig. 
\ref{FigOptimalErrorInstanton}.	
}
}
\label{FigOptimalNMInstanton}
\end{figure}
\begin{figure}[!ht]
\caption{
{
Behavior of the relative error obtained for the Airy function 
(dots) and for the instanton function (solid curve) versus 
$N$, for $F=k=3$ (a), 7 (b), 12 (c), and 20 (d).
}
}
\label{FigComparisonAiryInst}
\end{figure}
\begin{figure}[!ht]
\caption{
{Pictorial representation of the 
saddle networks and of the complex integration
paths involved in the evaluation of the Airy (a) and
of the instanton (b) functions.}
}
\label{FigCOmpAiryInst}
\end{figure}
\begin{figure}[!ht]
\caption{
{The same as in Fig. \ref{FigCOmpAiryInst}, but for the evaluation 
of the swallowtail diffraction catastrophe at the points across the Stokes set
given by the triplets $(x,y,z)=(0, \kappa^{3/2}, \kappa^2\times\,0.23012\ldots)$, with $\kappa>0$.
}
}
\label{FigSwPaths}
\end{figure}

\begin{figure}[!ht]
\caption{
{
The same as in Fig.~\ref{FigErrorAiryF16}, but for the swallowtail 
function evaluated across the Stokes set $(x,y,z)=(0, \kappa^{3/2}, 
\kappa^2\times\,0.23012\ldots)$
with $\kappa=2$.
}
}
\label{FigSWk2}
\end{figure}
\begin{figure}[!ht]
\caption{
{
Behaviors, as functions of $\kappa\in[2,4]$, of the relative errors 
for the swallowtail function evaluated across the Stokes set 
$(x,y,z)=(0, \kappa^{3/2}, \kappa^2\times\,0.23012\ldots)$ obtained 
via the 1st- (open circles) and the 2nd-level (dots) H-WT. As for the 
results presented in Fig. \ref{FigOptimalErrorAiry}, each point has 
been obtained by carrying out, for a given $k$, an exhaustive search 
for those values of the truncations $N$ and $(N,M)$ which minimize 
the corresponding 1st- and 2nd-level relative errors, respectively.
}
}
\label{FigOptimalErrorsw}
\end{figure}
\begin{figure}[!ht]
\caption{
{
Behavior, as a function of $\kappa$, of the values of
$N$ corresponding to the optimal setting for the 1st-level H-WT in 
the experiment of 
Fig. \ref{FigOptimalErrorsw}.}
}
\label{FigOptimalNsw}
\end{figure}
\begin{figure}[!ht]
\caption{
{
Behavior, as a function of $\kappa$, of the values of
$N$ (a) and $M$ (b) corresponding to the optimal setting for the 
2nd-level H-WT in the experiment of Fig. \ref{FigOptimalErrorsw}.}
}
\label{FigOptimalNMsw}
\end{figure}

\newpage\

\centerline{\psfig{figure=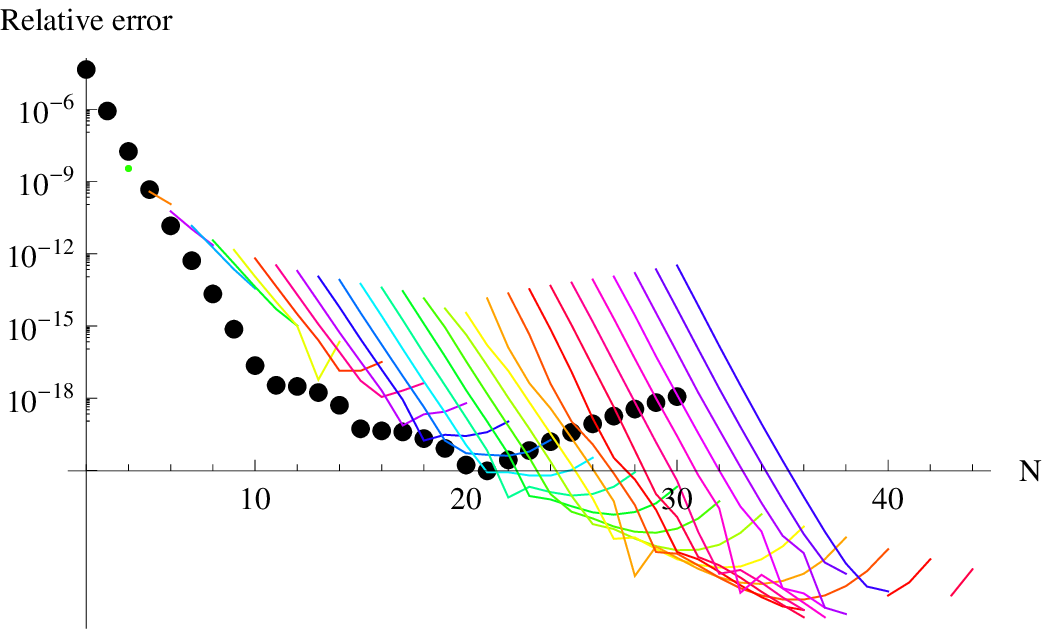,width=16cm,clip=,angle=0}}

\centerline{Fig. 1 - R. Borghi}

\newpage\

\centerline{\psfig{figure=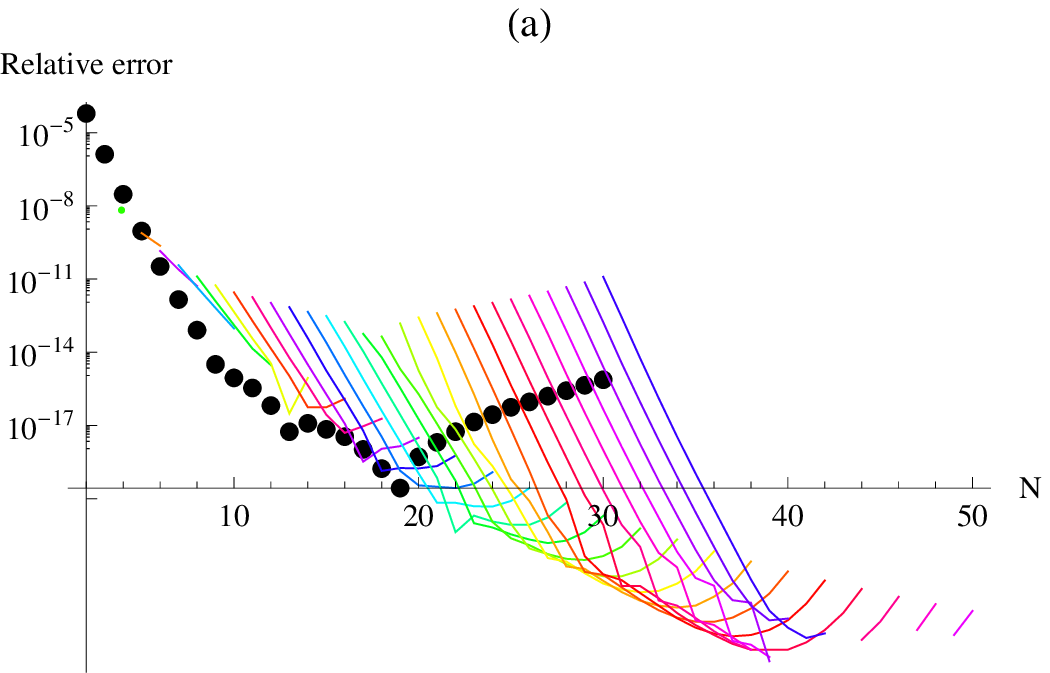,width=8cm,clip=,angle=0}}

\centerline{\psfig{figure=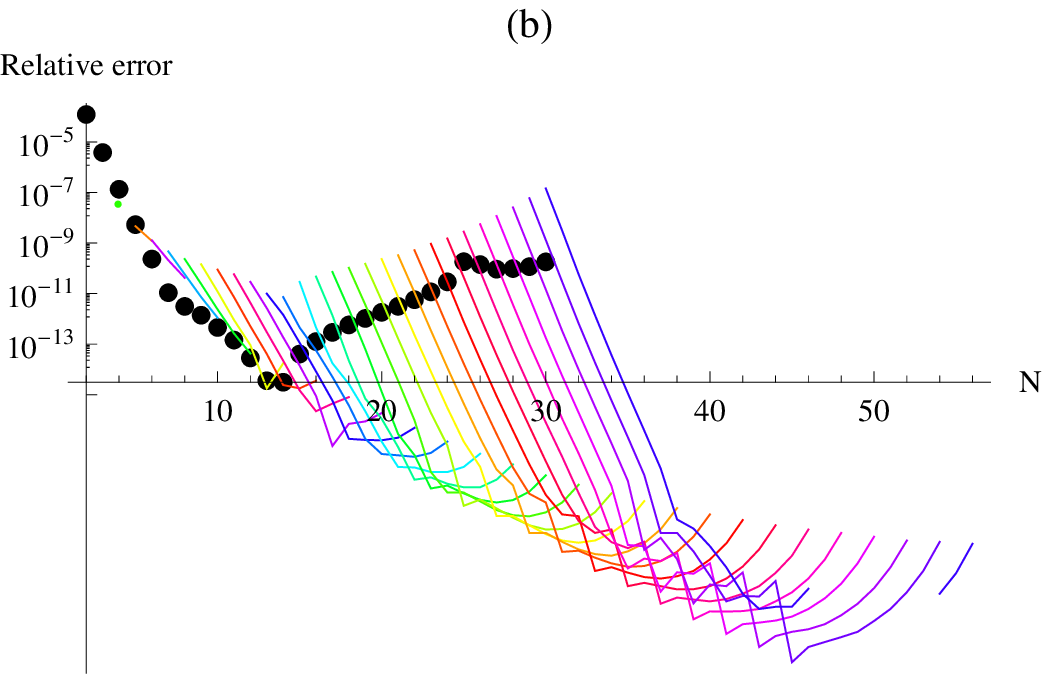,width=8cm,clip=,angle=0}}

\centerline{\psfig{figure=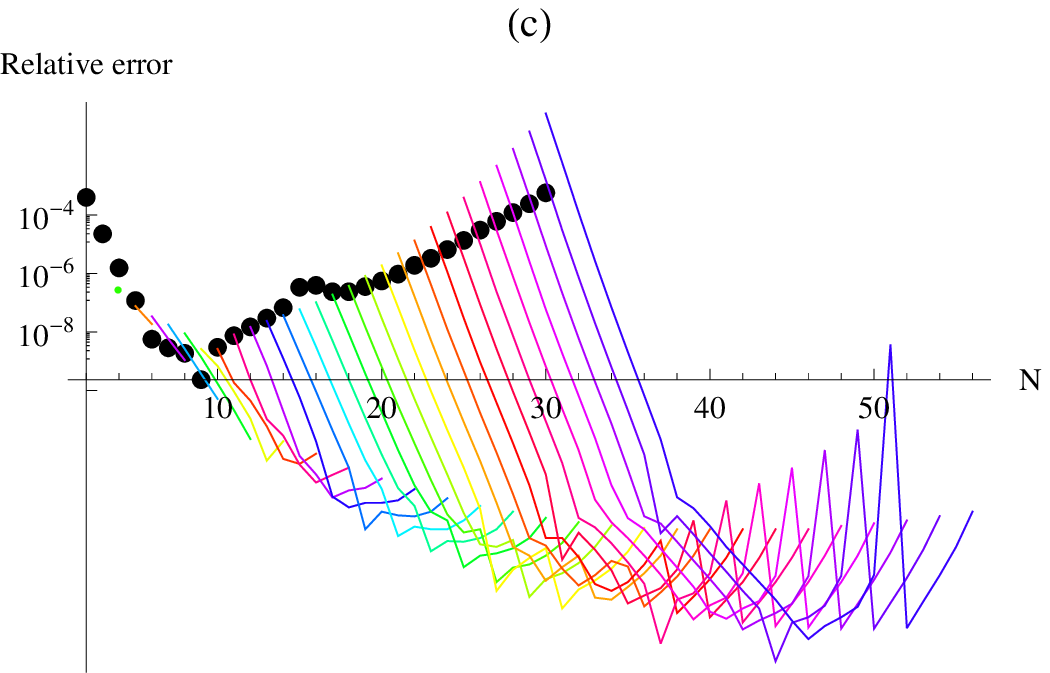,width=8cm,clip=,angle=0}}

\centerline{\psfig{figure=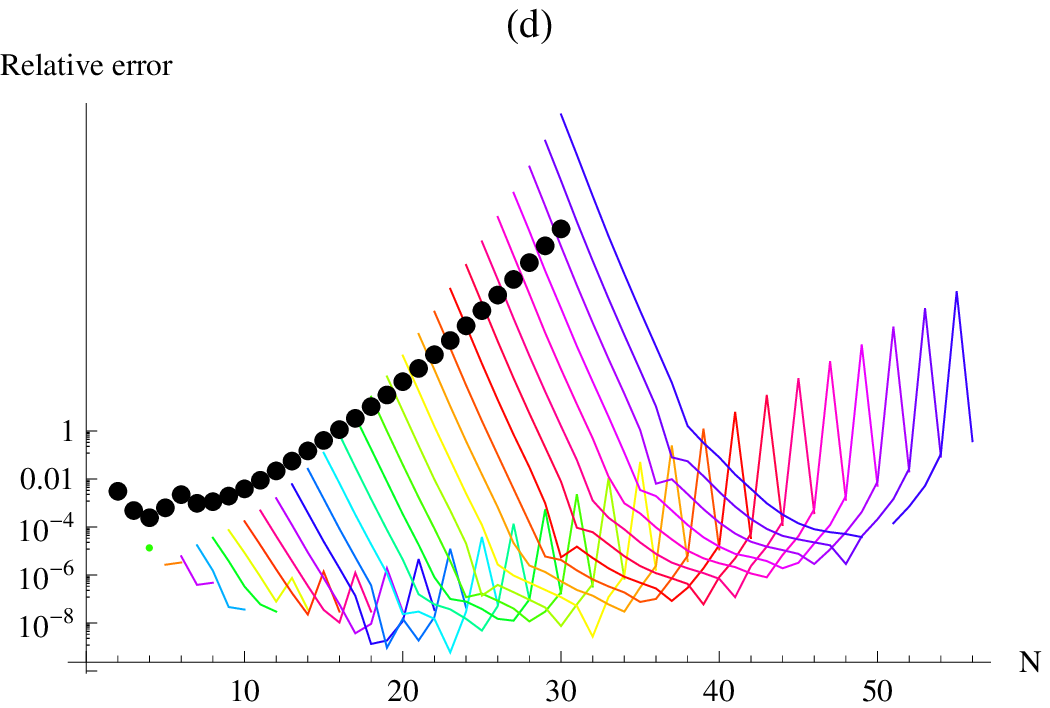,width=8cm,clip=,angle=0}}

\centerline{Fig. 2 - R. Borghi}

\newpage\

\centerline{\psfig{file=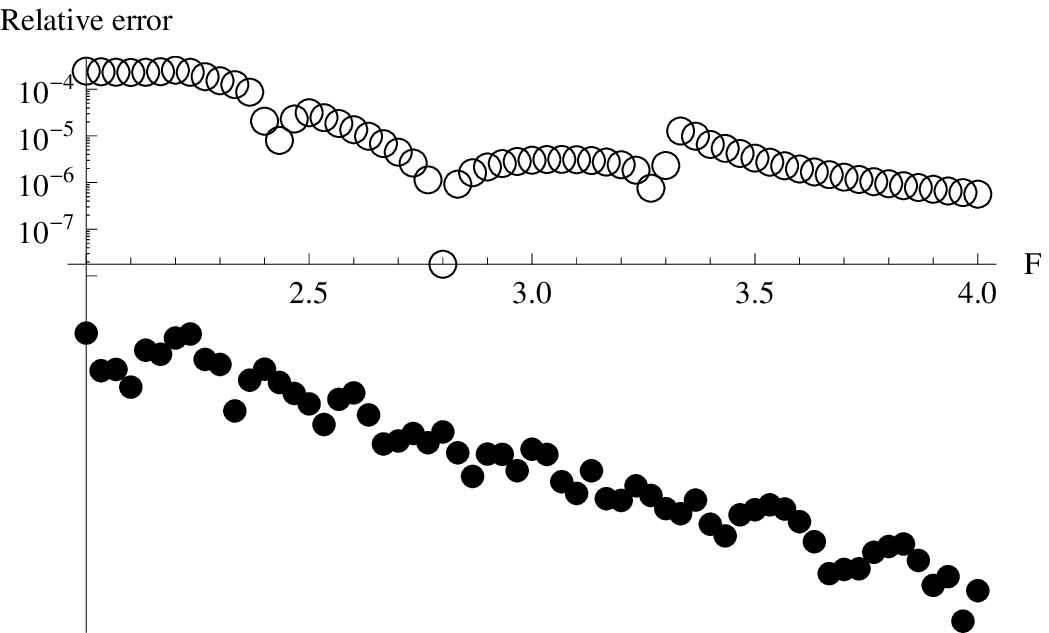,width=16cm,clip=,angle=0}}

\centerline{Fig. 3 - R. Borghi}

\newpage\

\centerline{\psfig{file=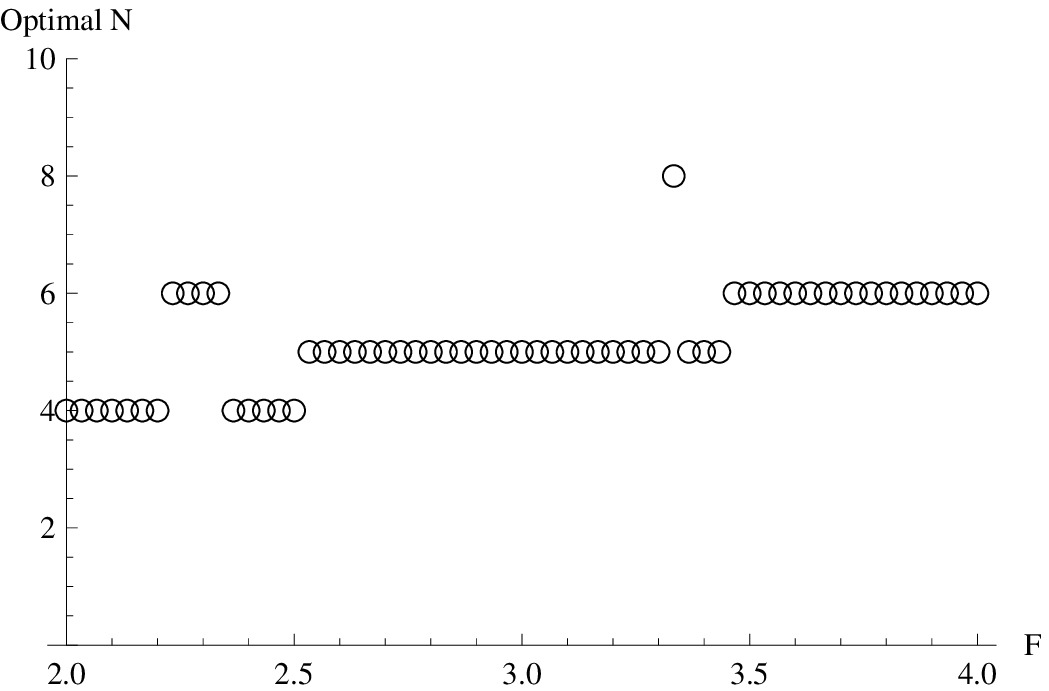,width=16cm,clip=,angle=0}}

\centerline{Fig. 4 - R. Borghi}

\newpage\

\centerline{\psfig{file=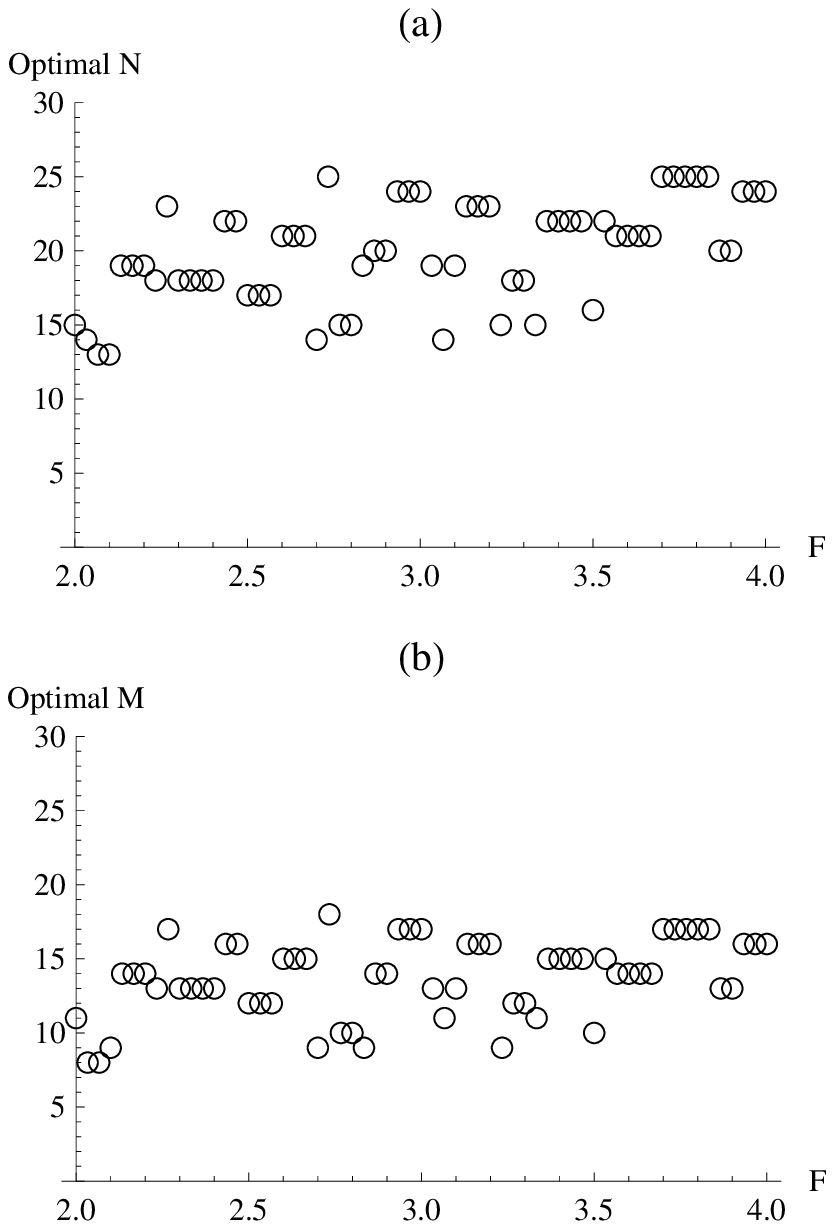,width=16cm,clip=,angle=0}}

\centerline{Fig. 5 - R. Borghi}

\newpage\

\centerline{\psfig{figure=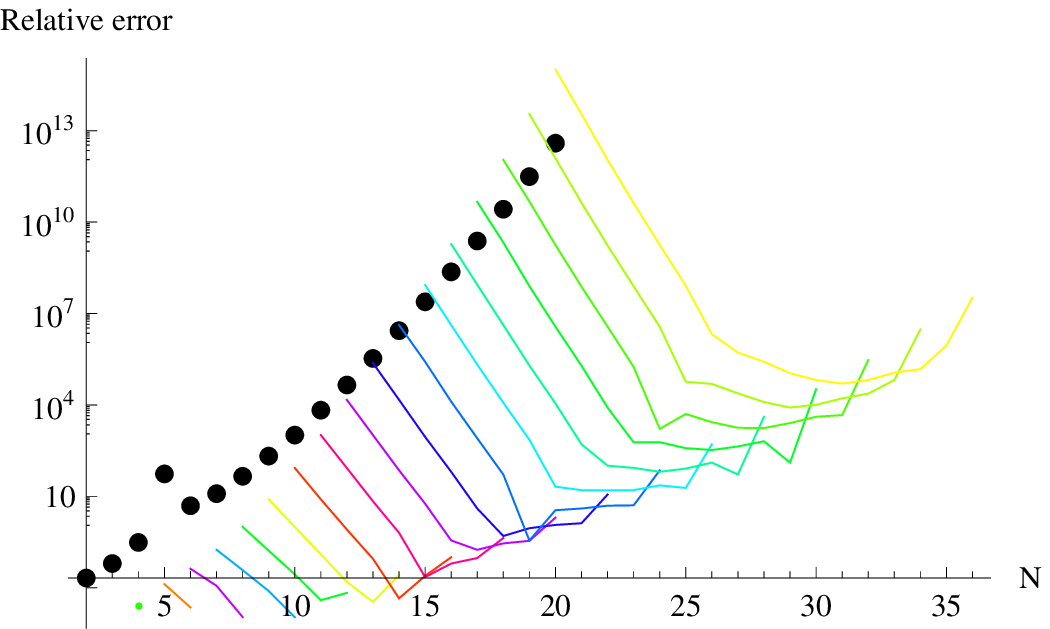,width=16cm,clip=,angle=0}}

\centerline{Fig. 6 - R. Borghi}

\newpage\

\centerline{\psfig{file=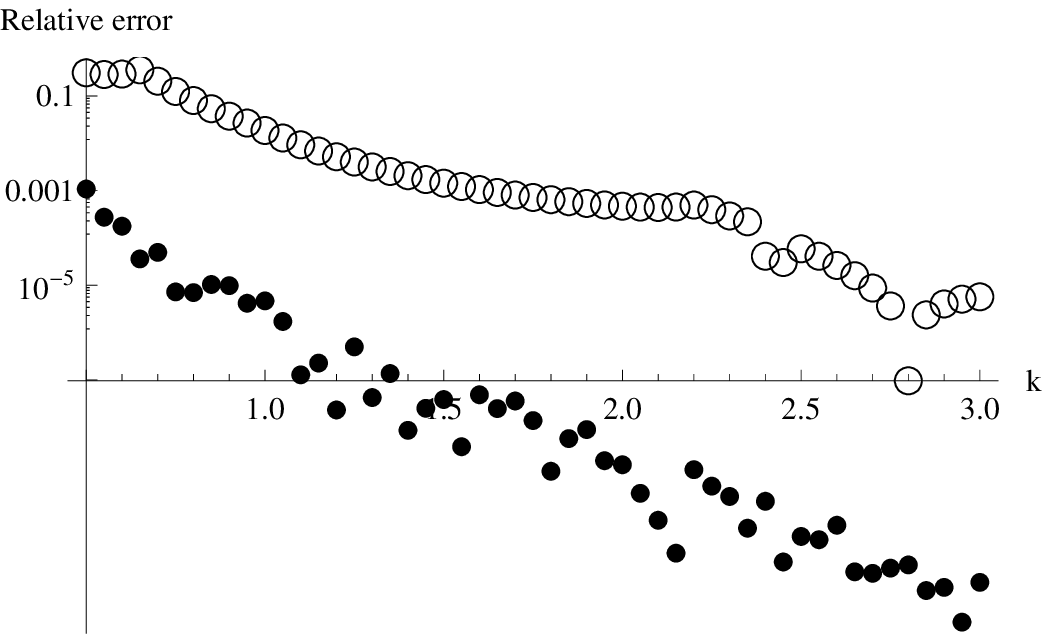,width=16cm,clip=,angle=0}}

\centerline{Fig. 7 - R. Borghi}

\newpage\

\centerline{\psfig{file=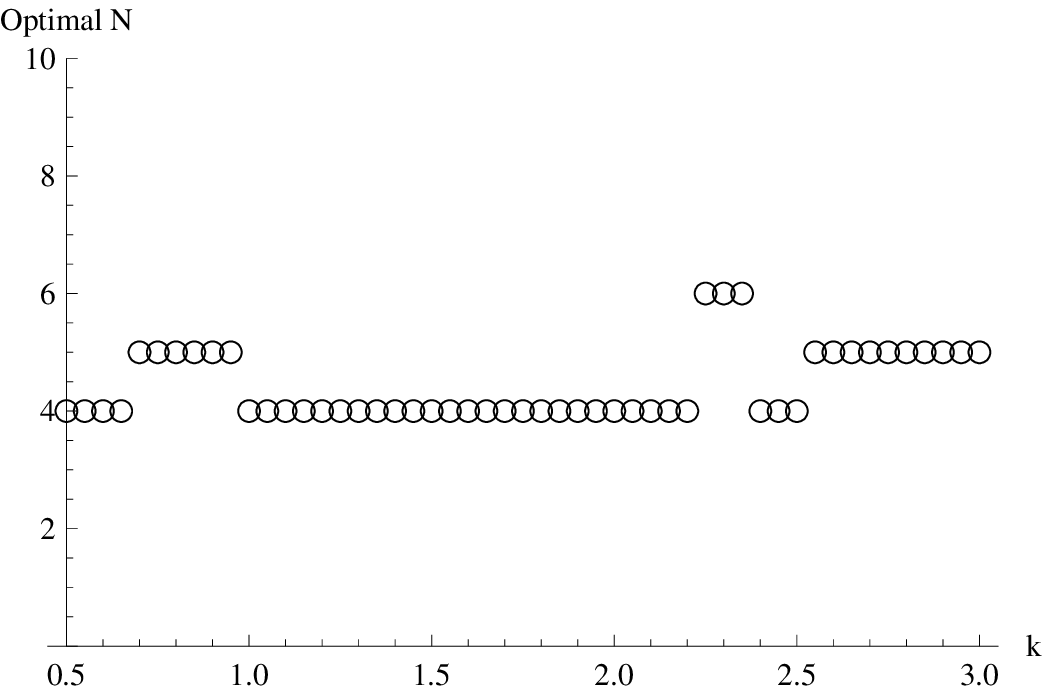,width=16cm,clip=,angle=0}}

\centerline{Fig. 8 - R. Borghi}

\newpage\

\centerline{\psfig{file=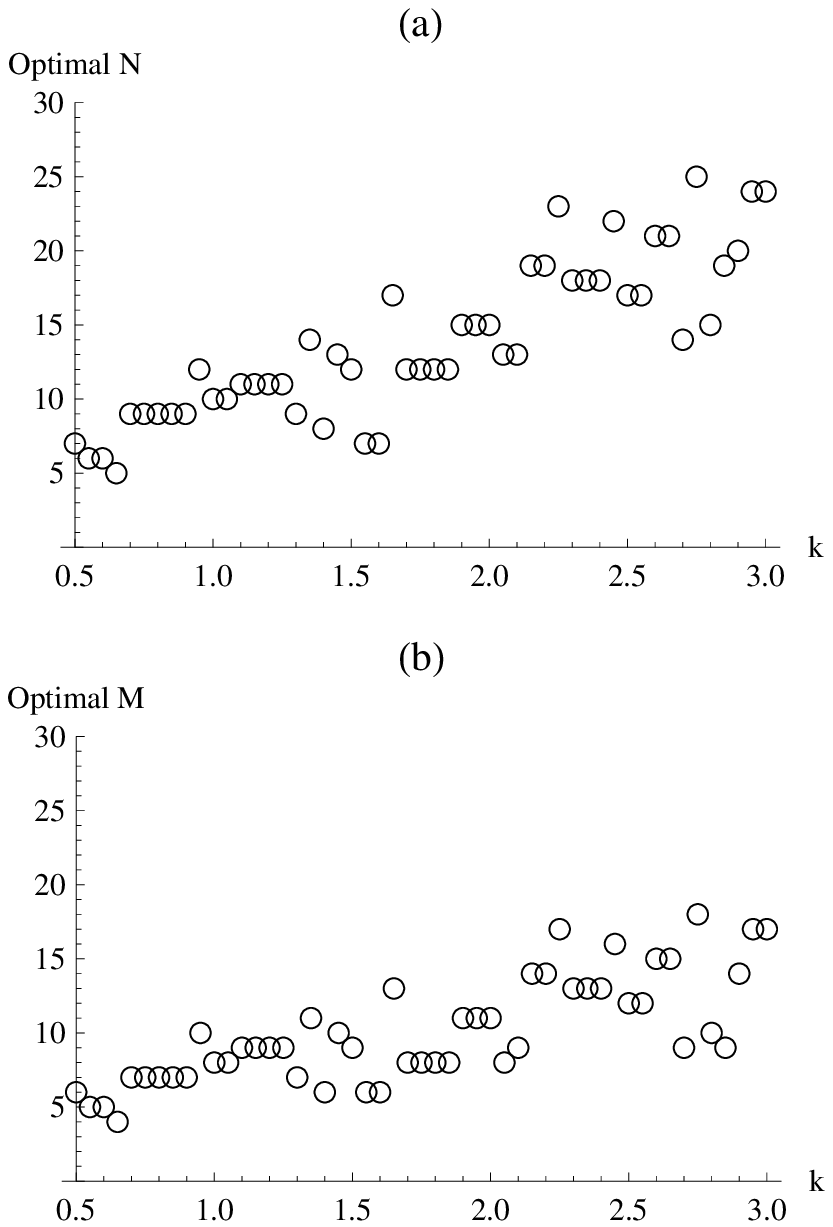,width=16cm,clip=,angle=0}}

\centerline{Fig. 9 - R. Borghi}

\newpage\

\centerline{\psfig{file=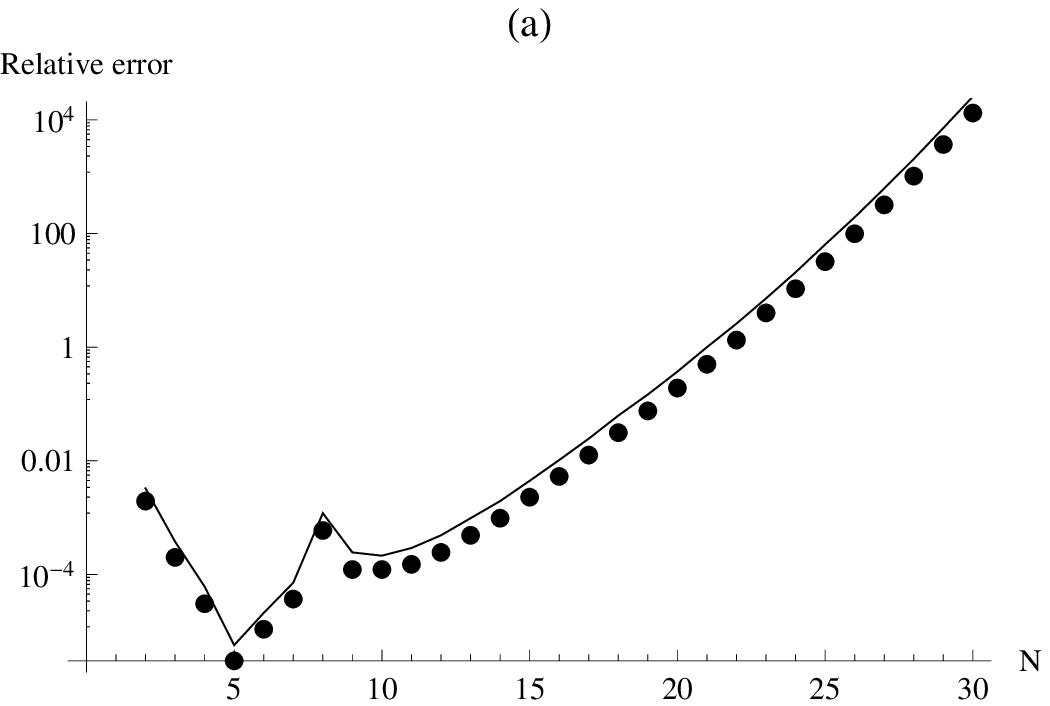,width=8cm,clip=,angle=0}}
\centerline{\psfig{file=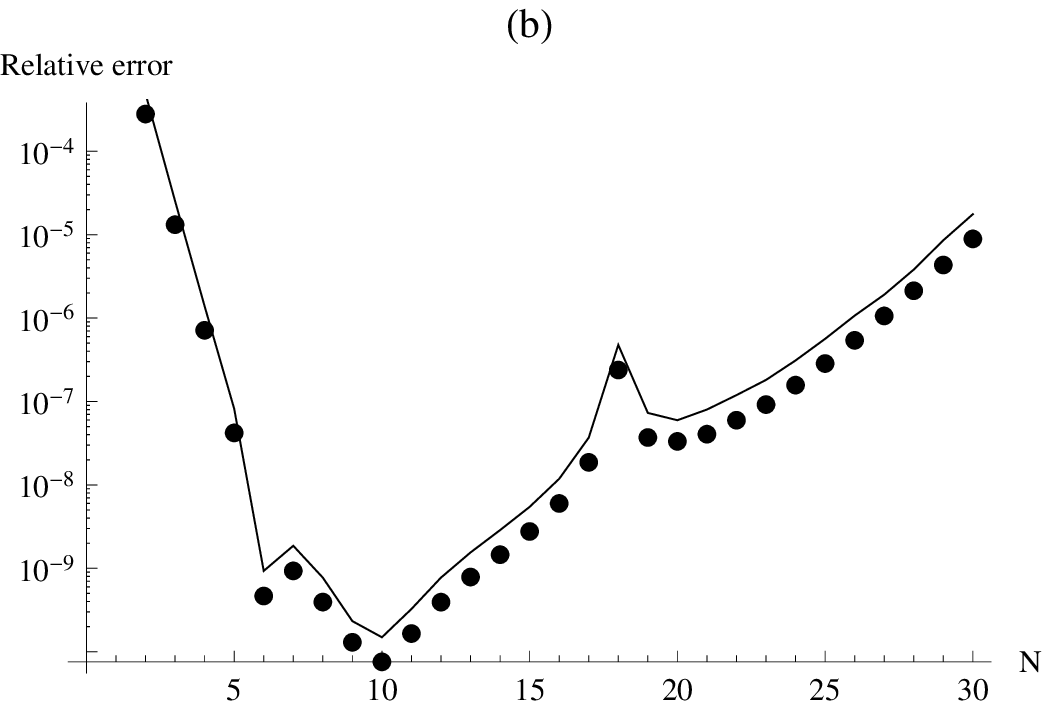,width=8cm,clip=,angle=0}}
\centerline{\psfig{file=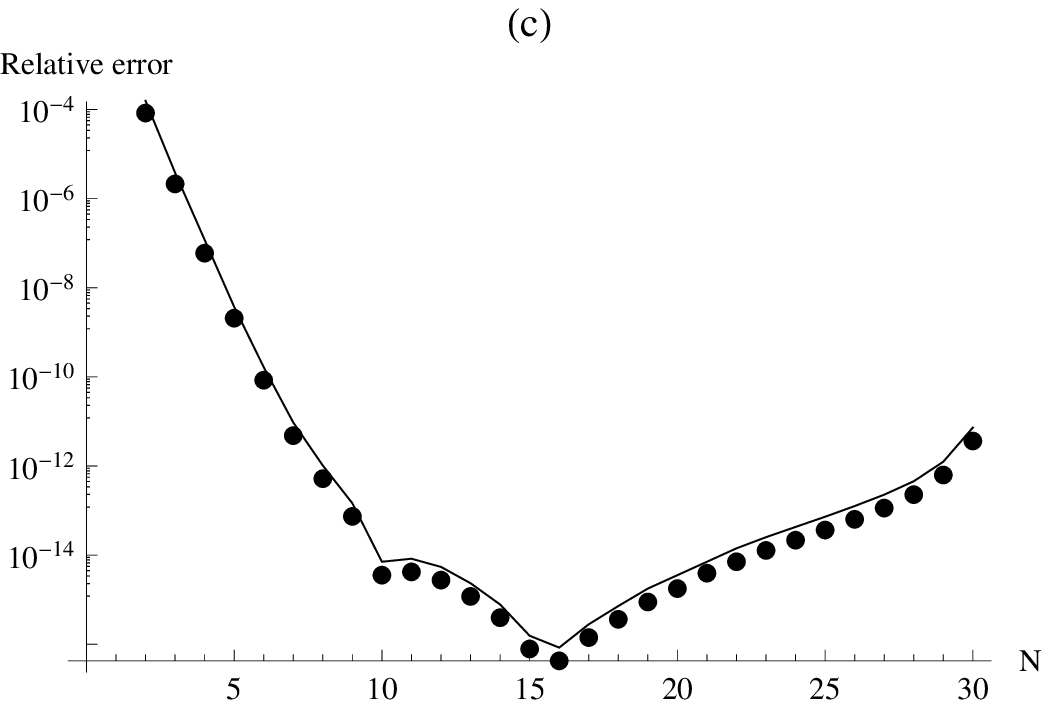,width=8cm,clip=,angle=0}}
\centerline{\psfig{file=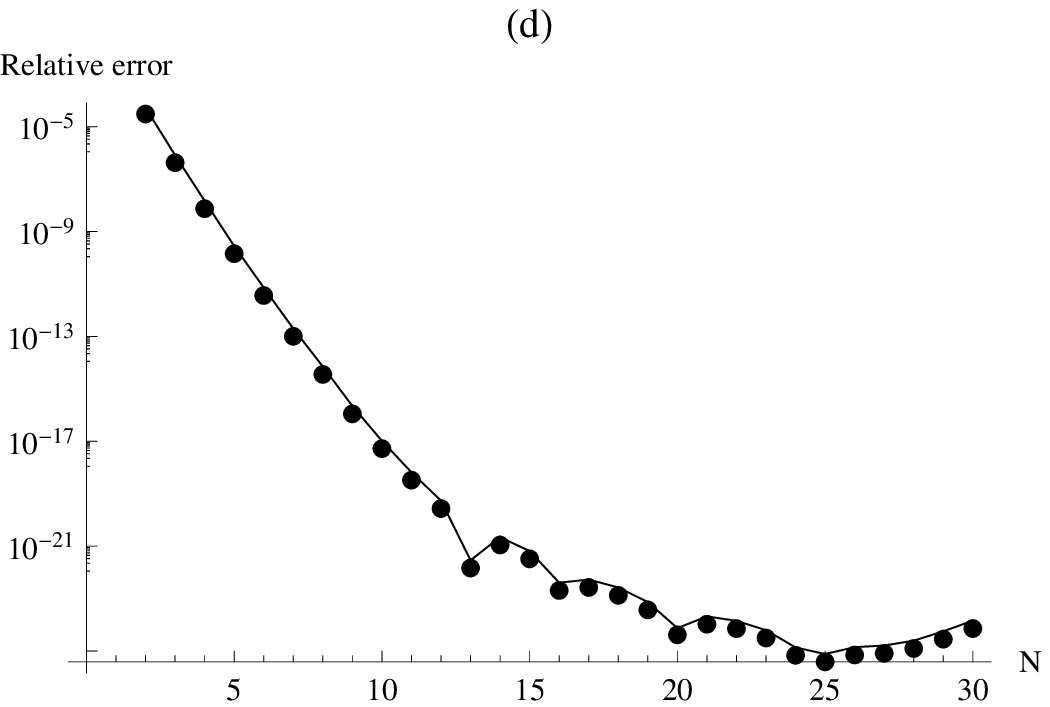,width=8cm,clip=,angle=0}}

\centerline{Fig. 10 - R. Borghi}

\newpage\

\centerline{\psfig{file=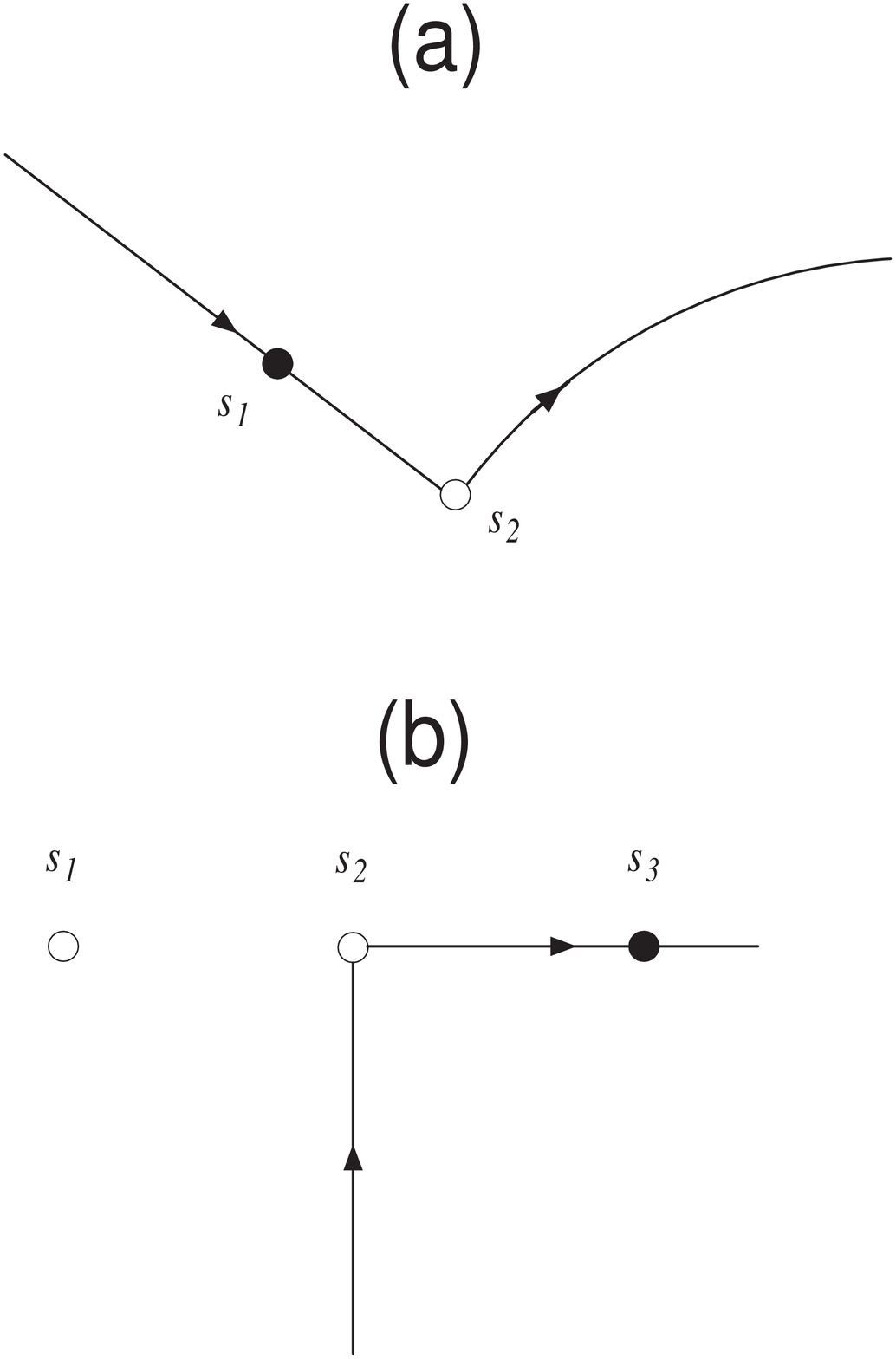,width=14cm,clip=,angle=0}}

\centerline{Fig. 11 - R. Borghi}

\newpage\

\centerline{\psfig{file=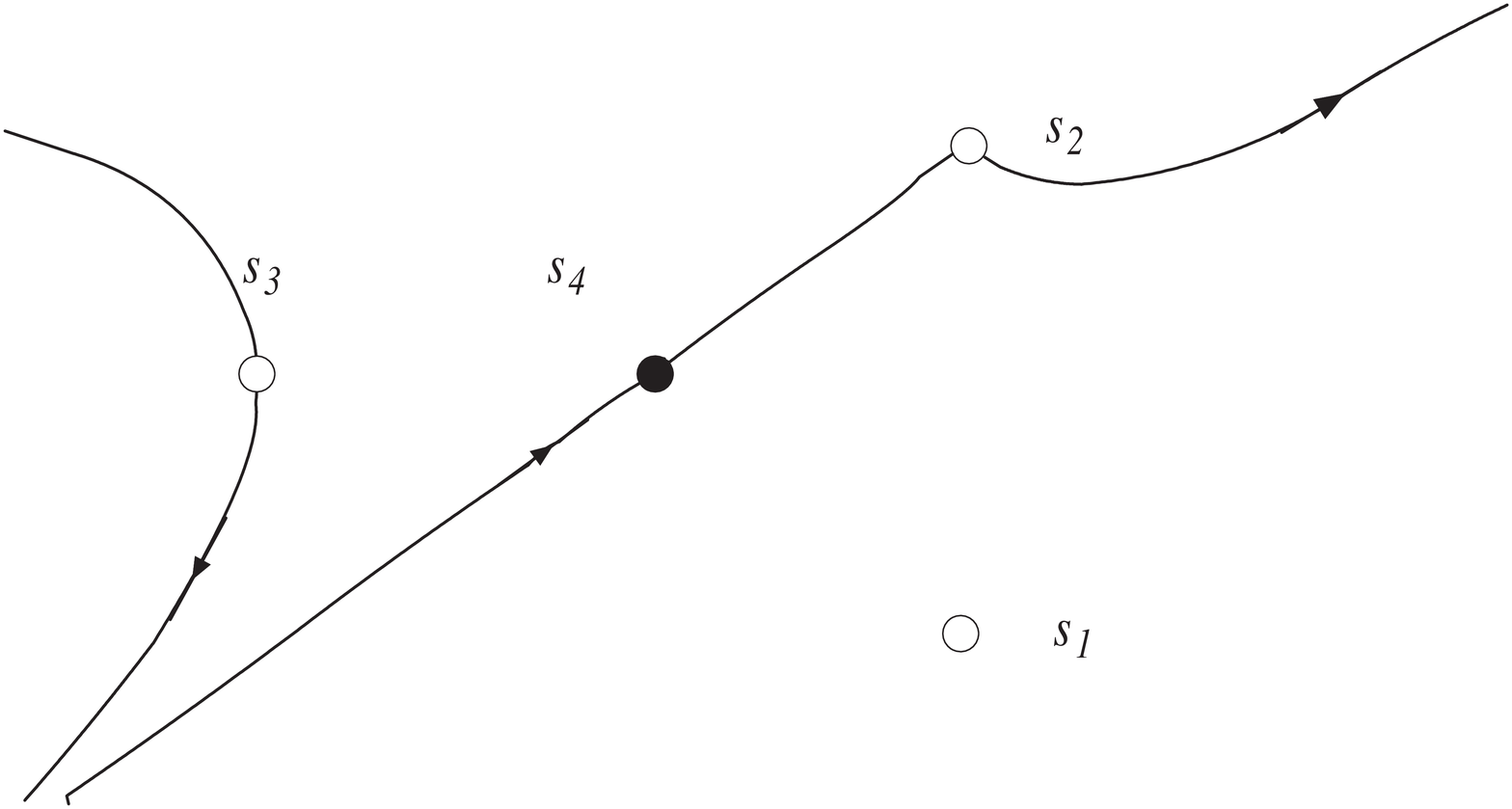,width=15cm,clip=,angle=0}}

\centerline{Fig. 12 - R. Borghi}

\newpage\

\centerline{\psfig{figure=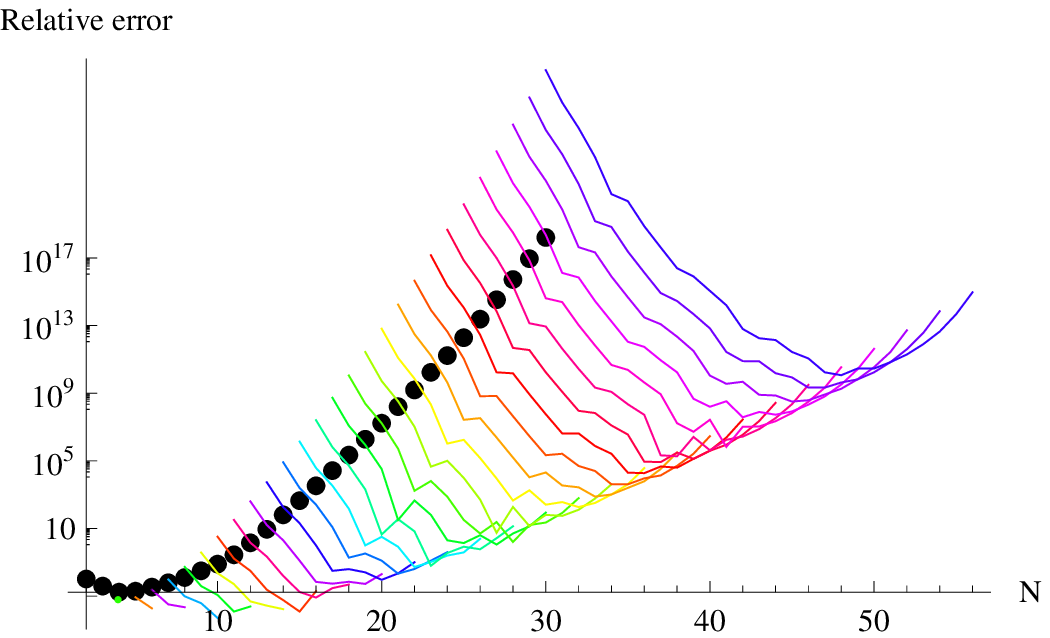,width=16cm,clip=,angle=0}}

\centerline{Fig. 13 - R. Borghi}

\newpage\

\centerline{\psfig{file=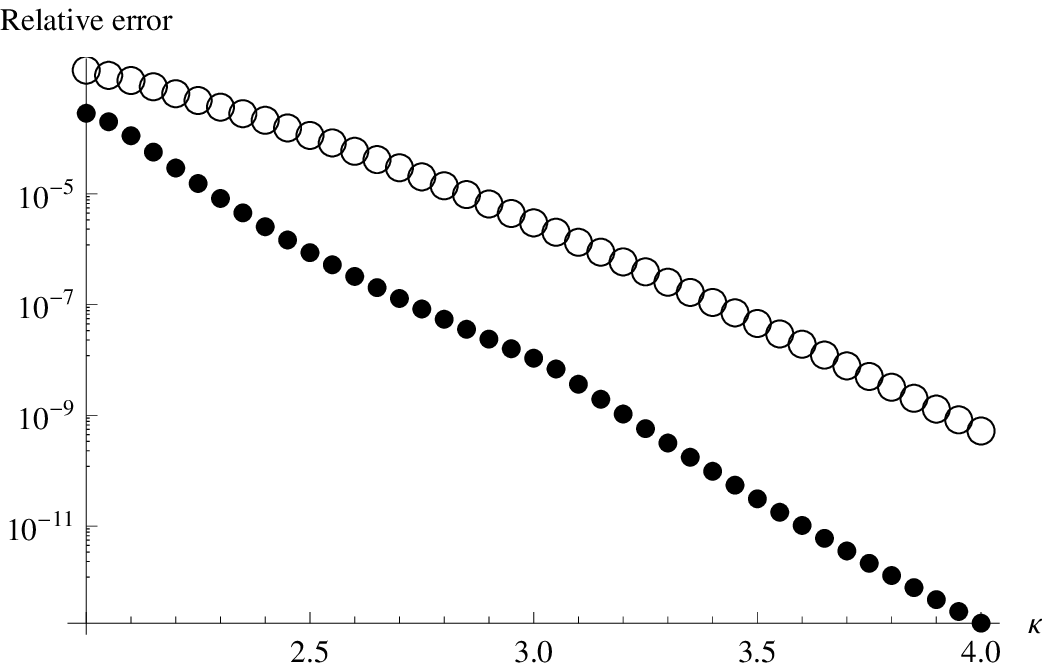,width=16cm,clip=,angle=0}}

\centerline{Fig. 14 - R. Borghi}

\newpage\

\centerline{\psfig{file=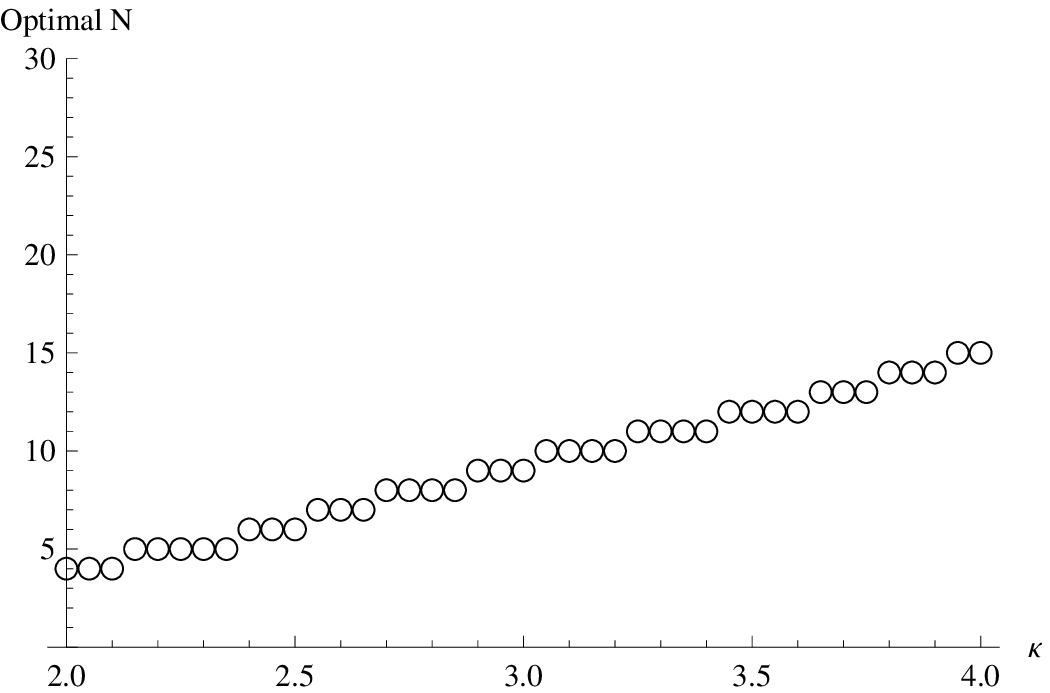,width=16cm,clip=,angle=0}}

\centerline{Fig. 15 - R. Borghi}

\newpage\

\centerline{\psfig{file=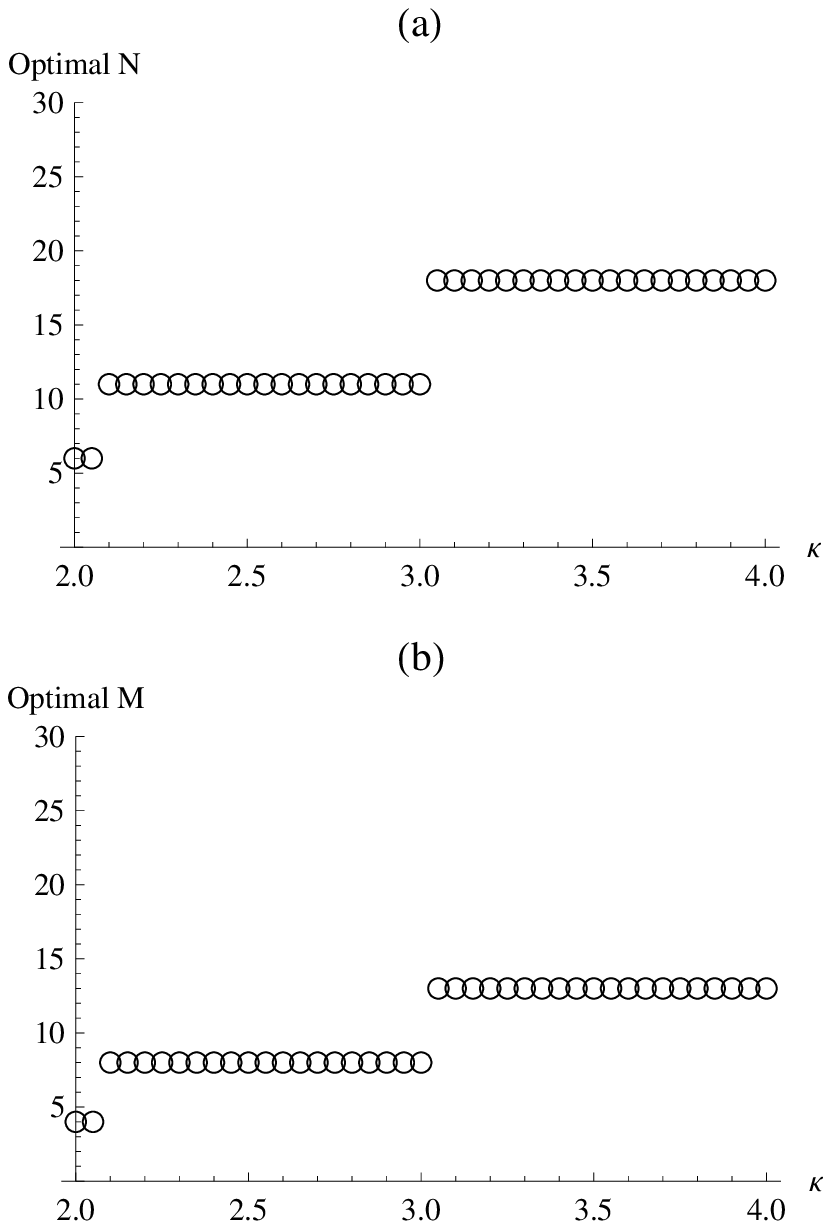,width=16cm,clip=,angle=0}}

\centerline{Fig. 16 - R. Borghi}


\begin{thebibliography}{00}
\bibitem{endnote31}
{The $\delta$- transformation was introduced in Ref. 
\cite{sidi} to derive
Pad\'e approximants for some hypergeometric functions, but only in 
Ref. \cite{wenigerCPR-89}
its usefulness as a powerful generalized resummation and convergence 
acceleration method was demonstrated.}
\bibitem{endnote30}
It could be proved that the above ``equivalence" between the Airy and the instanton functions keeps its validity also for the 2nd-level H-WT.
\end{thebibliography}
\end{document}